\documentclass[10pt,journal,compsoc]{IEEEtran}
\ifCLASSOPTIONcompsoc
  \usepackage[nocompress]{cite}
\else
  \usepackage{cite}
\fi
\ifCLASSINFOpdf
  \usepackage[pdftex]{graphicx}
\else
\fi
\usepackage{amsmath}
\ifCLASSOPTIONcompsoc
 \usepackage[caption=false,font=footnotesize,labelfont=sf,textfont=sf]{subfig}
\else
 \usepackage[caption=false,font=footnotesize]{subfig}
\fi

\usepackage{booktabs}

\begin{document}
%
\title{CTDGM: A Data Grouping Model Based on Cache Transaction for Unstructured Data Storage Systems}
%
%
%
%

\author{Dongjie~Zhu,~
        Haiwen~Du,~
        Yundong~Sun~
        and~Zhaoshuo~Tian,~\IEEEmembership{Member,~IEEE,}
\IEEEcompsocitemizethanks{\IEEEcompsocthanksitem D. Zhu and Y. Sun are with the School
of Computer Science and Technology, Harbin Institute of Technology, Weihai,
Shandong, China, 264209.
\IEEEcompsocthanksitem H. Du and Z. Tian are with the School
of Astronautics, Harbin Institute of Technology, Harbin, China, 150001.\protect\\
E-mail: duhaiwen@126.com
}
\thanks{Manuscript received April 27, 2020}
\thanks{(Corresponding author: Haiwen Du.)}}
%
%

\markboth{IEEE Transactions on Cloud Computing}%
{Shell \MakeLowercase{\textit{et al.}}: Bare Demo of IEEEtran.cls for Computer Society Journals}
%



\IEEEtitleabstractindextext{%
\begin{abstract}
  Cache prefetching technology has become the mainstream data access optimization strategy in the data centers. However, the rapidly increasing of unstructured data generates massive pairwise access relationships, which can result in a heavy computational burden for the existing prefetching model and lead to severe degradation in the performance of data access. We propose cache-transaction-based data grouping model (CTDGM) to solve the problems described above by optimizing the feature representation method and grouping efficiency. First, we provide the definition of the cache transaction and propose the method for extracting the cache transaction feature (CTF). Second, we design a data chunking algorithm based on CTF and spatiotemporal locality to optimize the relationship calculation efficiency. Third, we propose CTDGM by constructing a relation graph that groups data into independent groups according to the strength of the data access relation. Based on the results of the experiment, compared with the state-of-the-art methods, our algorithm achieves an average increase in the cache hit rate of 12\% on the MSR dataset with small cache size (0.001\% of all the data), which in turn reduces the number of data I/O accesses by 50\% when the cache size is less than 0.008\% of all the data.
\end{abstract}

}

\maketitle

\IEEEdisplaynontitleabstractindextext

%
\IEEEpeerreviewmaketitle

\IEEEraisesectionheading{\section{Introduction}\label{sec:introduction}}

%
%
%
%
\IEEEPARstart{T}{he} fast development of cloud computing and mobile computing has enabled massive amounts of unstructured data, such as images, documents, and videos, by mobile devices using the cloud platforms. Distributed storage systems, such as the Hadoop Distributed File System (HDFS)\cite{shvachko2010hadoop} and Ceph\cite{weil2006ceph}, have advantages in terms of scalability, concurrency, and availability. Therefore, these systems have become the mainstream storage solution for cloud platforms and are used by large Internet-based companies such as Facebook, Netflix, Yahoo, and Amazon\cite{bende2016dealing}. Until 2017, Spotify and Yahoo stored approximately 50\% of small files with a size of less than 1 MB, and the number of I/O operations on small files accounted for approximately 50\% of all I/O operations\cite{niazi2018size}. In addition, the rapid development of data science\cite{wang2017odds} has increased the pressure on small-sized unstructured data in distributed storage systems. However, the storage mechanism of the distributed storage system is mainly designed for storing large files, and accessing a large number of small-sized unstructured data causes the storage system generate a large number of I/O operations on disk (e.g., metadata access, data access), which substantially affects the access concurrency and response time of the data center\cite{chandrasekar2013novel,niazi2017hopsfs}.

Researchers have used data merging and cache prefetching strategies to improve the data access efficiency of storage systems and to address these problems. Among these approaches, the data merging strategy merges small-sized unstructured data that are strongly correlated into a big file and store it in a consecutive location, which reduces the number of I/O operations and improves the indexing efficiency of the distributed storage system via hierarchical indexing\cite{dong2012optimized}. The cache prefetching technology predicts and caches the data that may be accessed at the next moment by estimating the correlation to previously accessed data. The cache prefetching technology has been reported to effectively reduce the data access latency, thereby improving the concurrency of the storage system\cite{lin2008amp,kroeger1997exploring}. These strategies have produced good results and are widely used in storage efficiency optimization \cite{dong2010novel,zhu2018access,xiong2019small}, data deduplication \cite{fu2014accelerating,clements2009decentralized} and other fields. Data grouping is a strategy that merges related data into mutually exclusive groups and accesses them by group, which effectively reduces the disk I/O of the storage system. The accuracy and computational complexity of the correlation mining algorithm, which is the core component of the data grouping model, are the key factors to address when aiming to improve the data access efficiency of the storage system\cite{tamersoy2014guilt}, which has become a popular issue to study.

However, data in storage systems are saved as blocks of different sizes that are scattered throughout disks instead of in one continuous location. This type of storage is critical for a multi-core server, but on such servers, the order of the blocks in a file is not maintained. Therefore, block layer correlations are common semantic patterns in storage systems\cite{li2004c}. Additionally, a block layer I/O trace can provide detailed information about request queue operations and is collected by utilities such as blktrace; thus, block-level correlation mining is widely studied\cite{ge2018chewanalyzer,liao2018block}. Our study focuses on the data correlation in the block layer. In the present study, the term data refers to the continuously saved data in the block layer.

Data possess fewer valuable features for correlation mining. Most existing data correlation analysis methods use the block address and the access interval between data as features and estimate data correlations by defining the distance between data\cite{wildani2016can,yang2017mithril,liao2015prefetching,liao2015performing}. These correlation analysis methods determine the data features from related data, and such features are called relationship defined features (RDFs). Moreover, the distributed storage system runs in multi-appli-cation and multi-user environments, and the correlations between data change dynamically with time\cite{hankins2003data}. Therefore, an RDF is an unstable feature that requires a large amount of primary storage space to save pairwise relations between data, and it must be recalculated when access patterns change, resulting in substantial computational costs and memory consumption. Additionally, since the cache space is limited, the size of the data may affect the distribution of other data in the cache, which in turn affects the data correlations. However, existing methods for analyzing data relationships ignore the impact of the size of the data on data correlations.

We proposed a data correlation feature called the cache transaction feature (CTF). Unlike the RDF, a cache transaction reflects the cache distribution of a real storage system. A cache transaction consists of a set of data that appears together in the cache; therefore, the data size can be considered in the data correlation analysis, and the data access feature is represented as a vector of the cache transactions in which the data appear. Thus, the feature is determined by not only the pairwise access relation but also the data itself, which substantially reduces the memory consumption required for relationship storage.

Furthermore, we propose a data grouping model, CTDGM, based on a cache transaction. CTDGM aims to obtain accurate and closely related data groups by merging highly correlated data. In terms of computational efficiency, CTDGM divides the data grouping process into data chunking and group merging processes. The data chunking algorithm divides the data into data chunks according to access times, spatial locality and CTF. The group merging process is performed on highly related data chunks according to the cache transaction, which potentially improves the mining range and the inner relation strength of the group to provide a more accurate relation mining result. Moreover, the complexity and number of calculations will be substantially reduced by the two-layer relation mining method. The main contributions of our study are listed below.

\textbf{1) A more accurate representation of the relationships between data:} For RDF, the size of the data is often ignored. The CTF incorporates the size of the data into the data access feature. Moreover, it presents the access feature of the data as a vector, which makes the value easy to calculate. Therefore, it is suitable for data relation mining in the data center and expresses more access features of the data.

\textbf{2) Calculate the data correlation without a spatial locality limitation:} CTDGM improves the range of relation mining using the two-layer mining algorithm, which reduces the negative impact of block address mining range on the computing efficiency and thus potentially increases the mining accuracy.

\textbf{3) Ensure that data with strong correlations are accurately mined:} The independence of the data group requires that data belong to only one group. CTDGM prioritizes the merging of data highly correlated to ensure only the most closely related data are merged and thus ensuring a high group inner relation strength.

The CTDGM model works on block I/O level with low computation overhead. It can also be used to group and prefetch related data on proxy workloads. Through cross-validation of widely used public datasets, our data grouping model significantly improved the cache hit rate, reduced the number of the disk I/Os, and reduced the data access latency.

In Section 2, we introduced the background of and motivation for the study. We describe the definition and method used to calculate the cache transaction in Section 3. In Section 4, we provide a detailed description of the data grouping algorithm. Section 5 presents the evaluation methodology and results. Related studies are discussed in Section 6. Finally, we provide some conclusions in Section 7.


\section{BACKGROUND AND MOTIVATION}


\subsection{Data access feature extraction}
As the basis of data access correlation mining, the method for extracting access features is a popular topic in the field of storage research. Most existing data correlation mining algorithms are based on an RDF or Ordinal Feature (OF). We provide a detailed explanation of the RDF and the OF below.

\textbf{Ordinal Feature (OF):} An OF indicates that the data are treated as separate units and are uniquely marked as independent features. Therefore, an OF is a suitable feature for the calculation of temporal locality. Researchers\cite{marascu2005mining,mishra2012discovery} use the OF-based frequent itemset mining algorithm to analyze the correlations of Web resources and prefetch files with high relevance. However, OF does not represent the spatial locality, which reflects the potential correlation between data\cite{jalaparti2018netco}. When an OF is applied to NN-based data prefetching strategies, it is represented as a one-hot encoded vector that requires extensive storage resources\cite{patra2010file}. Therefore, in recent years, researchers have attempted to mine data correlations based on RDF.

\textbf{Relation Defined Feature (RDF):} An RDF indicates that the access feature of the data is defined by the related data. Compared with an OF, an RDF is represented in pairwise relations between data, and it introduces logical block addresses as a factor that influences the data correlations, which potentially improves the relationship mining accuracy. Most RDF-based correlation mining models define the distance between data and regard it as the closeness of the data \cite{dong2010novel}. However, compared to an OF, the RDF is calculated by determining the distance from other data, causing the data to lose their independent features. Additionally, the pairwise relations among massive amounts of data must be pre-calculated and stored, which requires extensive computing and storage resources. 

\begin{figure}
\centering
\includegraphics[width=3.2in]{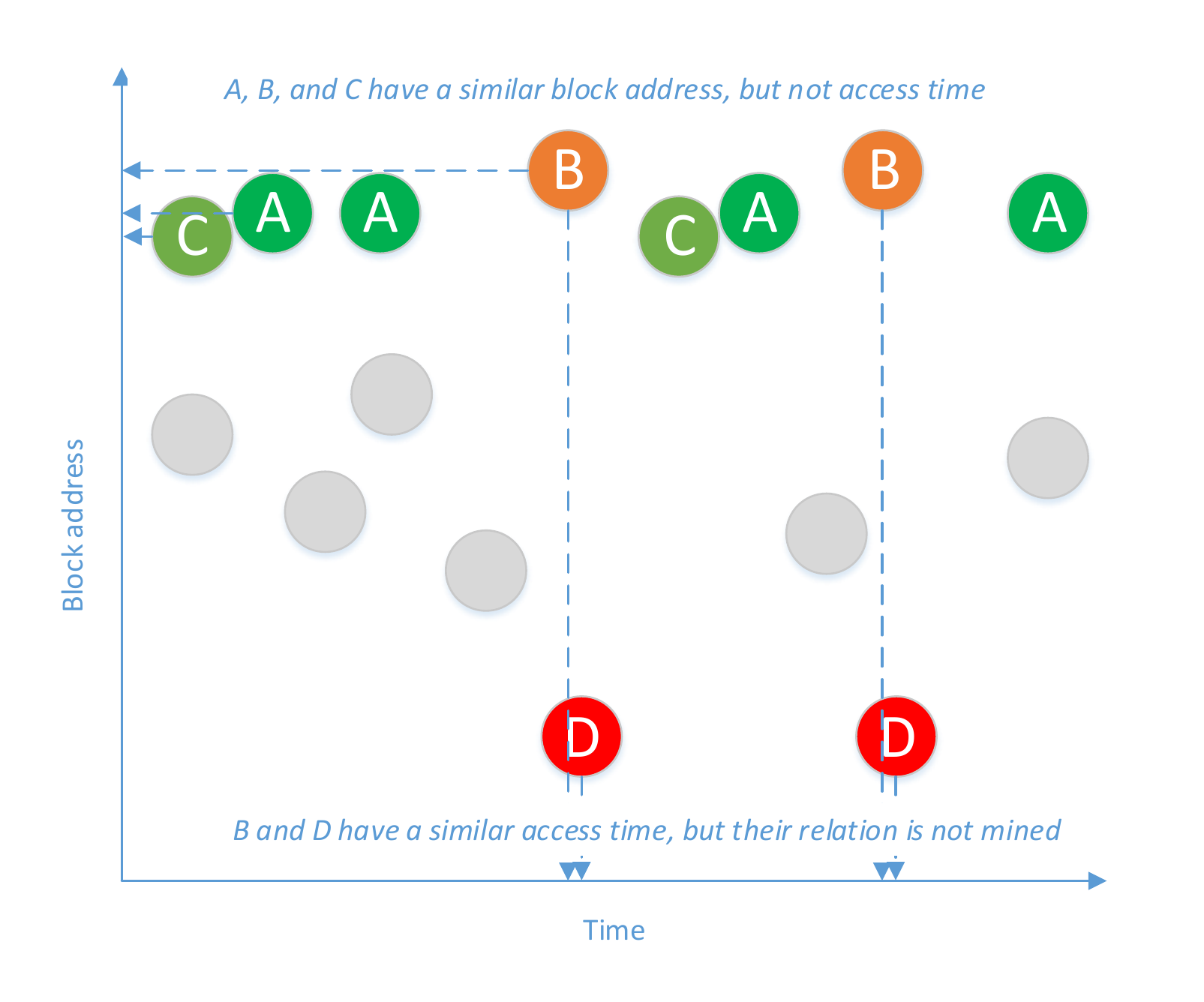}
\caption{An RDF-based relationship mining algorithm thoroughly exploits the relationships among A, B, and C. However, the relationship between B and D is difficult to mine.}
\label{figure1}
\end{figure}

Several RDF-based relation mining models perform the relational calculation only on data with close block addresses to ensure computational accuracy. Otherwise, the method may generate $n^2$ relations. These methods ignore data with large block address distances and have strong temporal locality, as shown in Fig. \ref{figure1}.

\begin{figure}[h]
  \centering
  \includegraphics[width=3.5in]{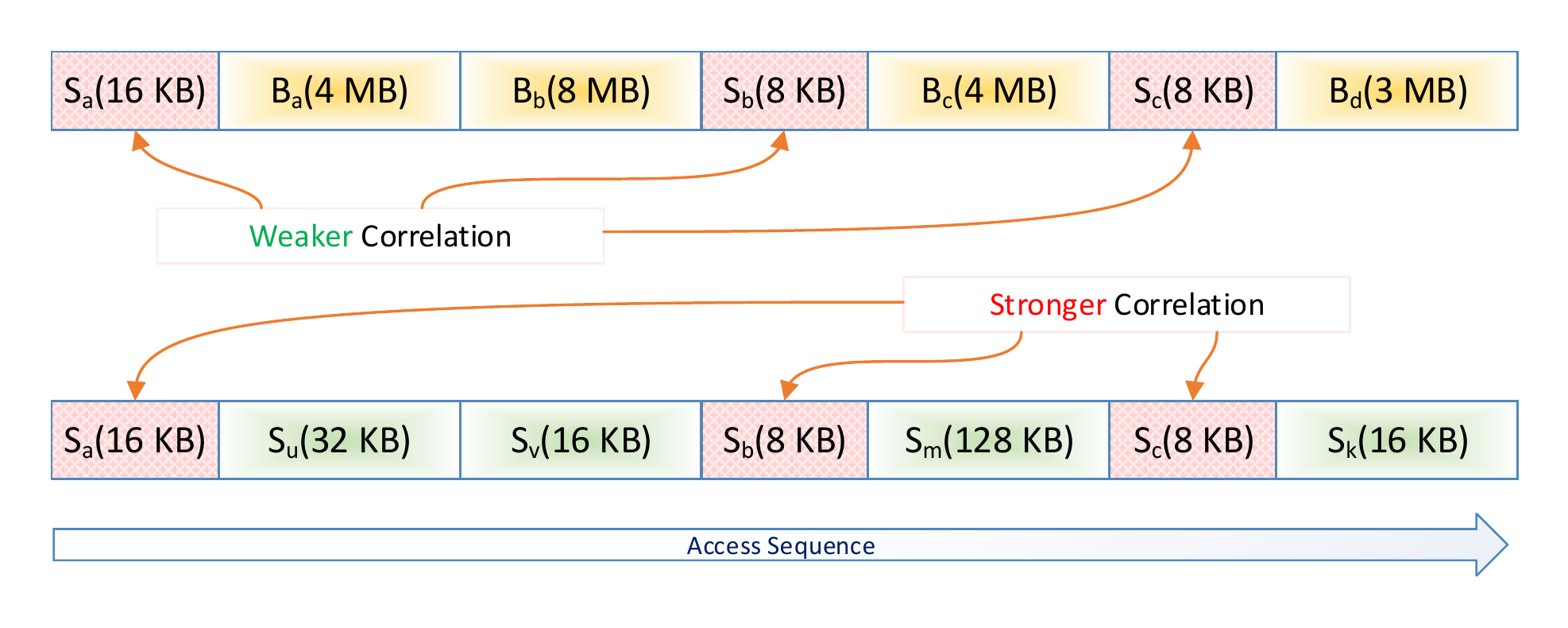}
  \caption{In the case of a limited cache size, a higher correlation is observed between Sa, Sb, and Sc in the lower access sequence, since they are more likely to appear in the cache at the same time.}
  \label{figure2}
  \end{figure}

Additionally, we identified the problem in which both the OF and RDF are unable to perceive the size of the data. In a real storage environment, the storage system concurrently receives access requests for both large data and small data. As shown in Fig. \ref{figure2}, when large data are accessed, the relevance of the data in the cache around the accessed data may also be affected because they occupy most of the cache space.

Therefore, the data size is a notable factor that the relationship mining algorithm should consider. The data near the larger data in the access sequence should be less relevant. We propose a feature representation based on cache transactions, which introduces the size of the data to the correlation mining process to solve this problem; thus, the degree of access relation between data is dynamically adjusted according to the size of the accessed data.

\begin{table*}[!t]
  \renewcommand{\arraystretch}{1.1}
  \caption{Comparation of prefetching approaches.}
  \label{table_0}
  \centering
    \setlength{\tabcolsep}{2.3mm}{ 
    \begin{tabular}{cccccc}
    \toprule
    Algorithm &Feature used&	Space overhead&	Time overhead	&Wasted Prefetches &Method	\\
    \midrule
    AMP\cite{gill2007amp}	&OF	&Low&	Low	&Moderate&Adaptive Asynchronous Prefetch	\\
    NN\cite{patra2010file}	&OF	&Low&	High&	High&	Neural Network\\
    MITHRIL\cite{yang2017mithril}	&RDF &High&	Moderate&	Moderate &	Distance between inverted data access index\\
    Tombolo\cite{yang2016tombolo}	&OF	&High	&Moderate& Moderate& Directed graph\\
    C-MINER\cite{li2004c}	&OF	&Moderate&	High&	Moderate&	Frequent pattern\\
    Wildani’s\cite{wildani2016can}	&RDF&High&	Moderate&Low &	Relation graph\\
    CTDGM	&CTF	&Moderate	&Moderate	&	Low &Cache transactions and Complete subgraph\\
    \bottomrule
    \end{tabular}}
    \end{table*}

\subsection{Data locality statistics}

\begin{figure}
  \centering
  \includegraphics[width=3.2in,trim=25 10 10 10,clip]{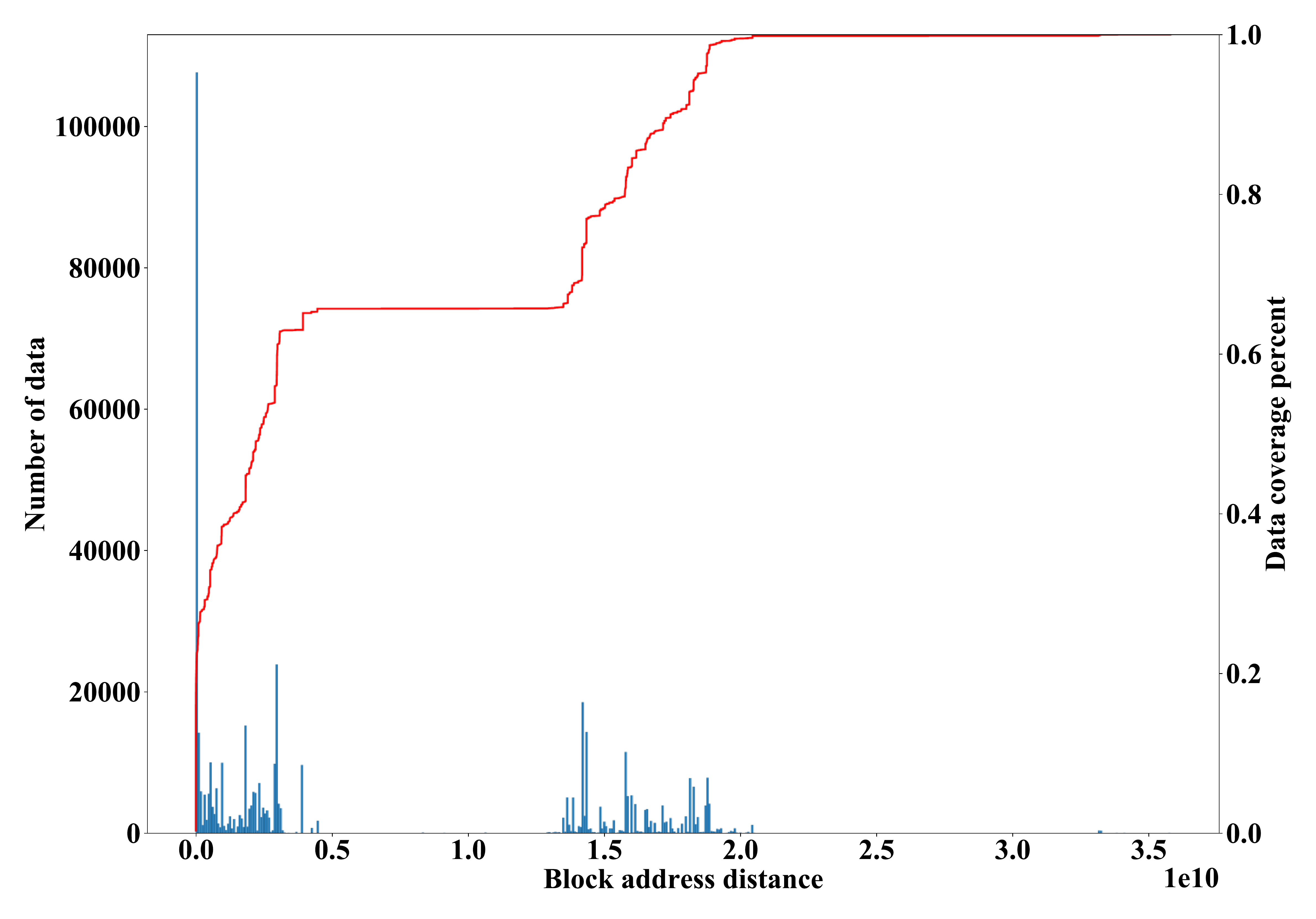}%
  \label{fig2_first_case}
  
  \caption{The blue line shows the distribution of data with a strong correlation in terms of the block address distance and the red line is the CDF of the blue line; we consider two data to be related when they are always accessed sequentially. Approximately 60\% of the data with strong correlations exist in close block addresses, but a large number of data with large distant block addresses (i.e., data with distances greater than 1.4e10) still exist.}
  \label{fig_sim_a}
  \end{figure}

In a distributed storage system, data access is characterized by spatio-temporal locality. Most studies use this feature to perform relation calculations and mine the access relationships among data\cite{liao2015prefetching}. However, since a large amount of data is stored in the storage system, the analysis of the pairwise relationships generated by the data may consume a large amount of computing resources. Therefore, most studies use the spatial locality of data as a range limitation of the access relation mining model. However, by analyzing the block address distance of the data with the associated relationship in the trace, temporal locality is indeed observed between the data with large block address distances, as shown in Fig. \ref{fig_sim_a}. As a result, since this method ignores a large number of relationships, the improvement in efficiency occurs at the expense of accuracy. 
  
\begin{figure}
    \centering
    \includegraphics[width=3.0in,trim=25 10 10 10,clip]{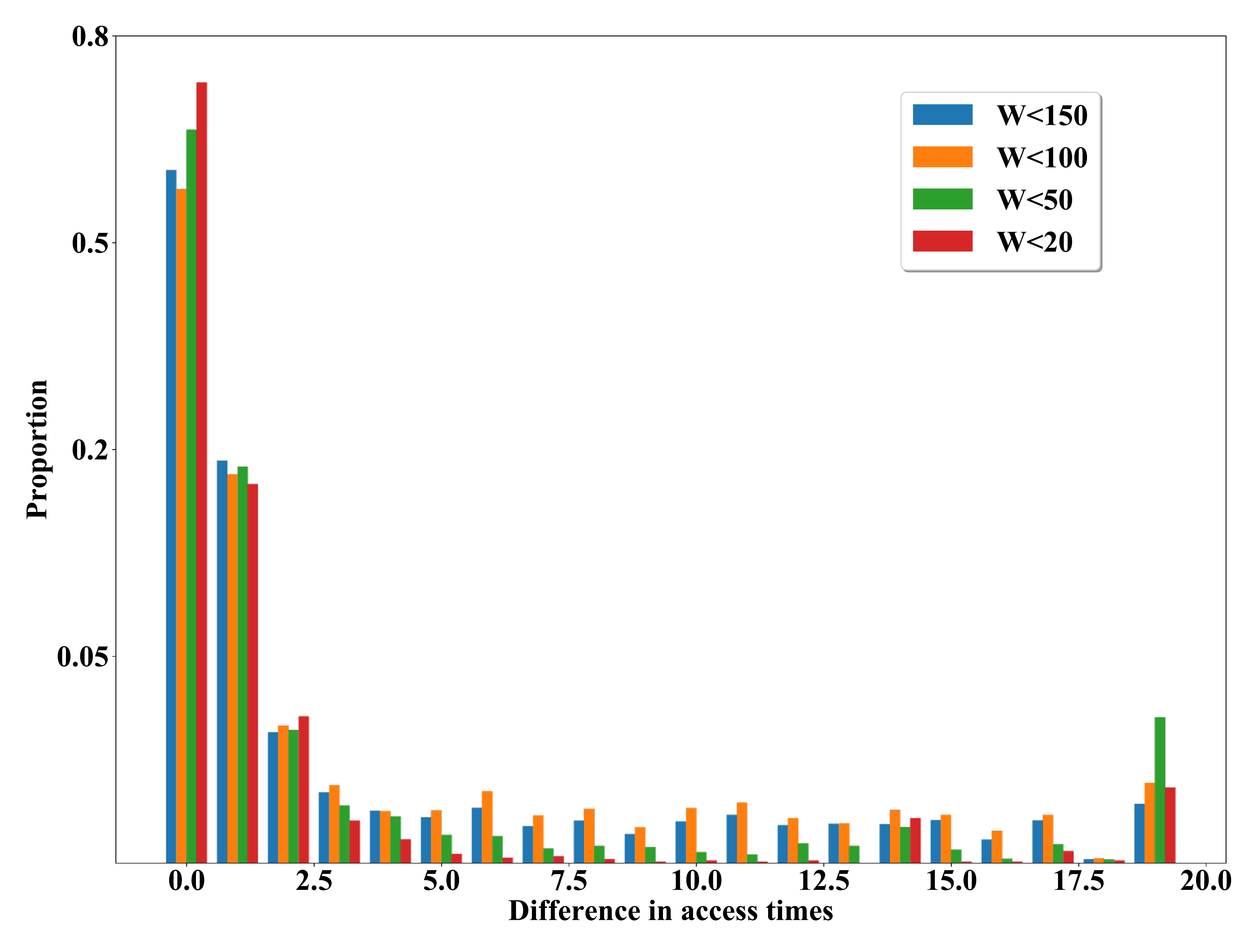}%
    \label{fig2_second_case}
    \caption{We use 4 different relation strength limits $W<20$, $W<50$, $W<100$ and $W<150$ to verify the effect of the difference in access times on data correlation. Only approximately 70\% of the data with access relevance have the same number of access times, and the proportion changes little under different relation strength limits.}
    \label{fig_sim_b}
    \end{figure}

  Researchers proposed Mithril \cite{yang2017mithril} for mining the access relation of data with large block address distances, based on the idea that most related data have similar access times. Mithril mines the access relations among the data with the same access times. However, we calculated the difference in access times to the data with access relevance, as shown in Fig. \ref{fig_sim_b}. Approximately 70\% of the data with access relevance have the same access times, and the proportion changes little under different relation strength limits. Therefore, the exclusive use of the access time of data for limiting the range of relation mining is inappropriate. Among the factors listed above, the relation strength $W$ is defined as the average access interval between data, as shown in Equation (1).

\begin{equation}
  W_{x,y}=\frac{\sum_{i=1}^{A_x}\textbf{min}\left ( \left | seq_{xi}-seq_{yj} \right | \right )}{A_x},j\in A_{y}
  \end{equation}

  For data $x$, $A_x$ is total number of accesses to data $x$, $seq_{xi}$ is the $i^{th}$ access sequence number of data $x$.

For closely related data, an analysis of the correlation of the data with similar access times and similar block addresses is efficient. However, as discussed above, this approach may ignore many relationships between blocks that are farther part in block address. Therefore, this method should be used only as a preliminary relationship mining method. A correlation mining algorithm that can mine data access relationships for data that are farther apart in block address is needed as a supplement.
  
\subsection{Prefetching algorithms}

\begin{figure}[h]
\centering
\includegraphics[width=3in]{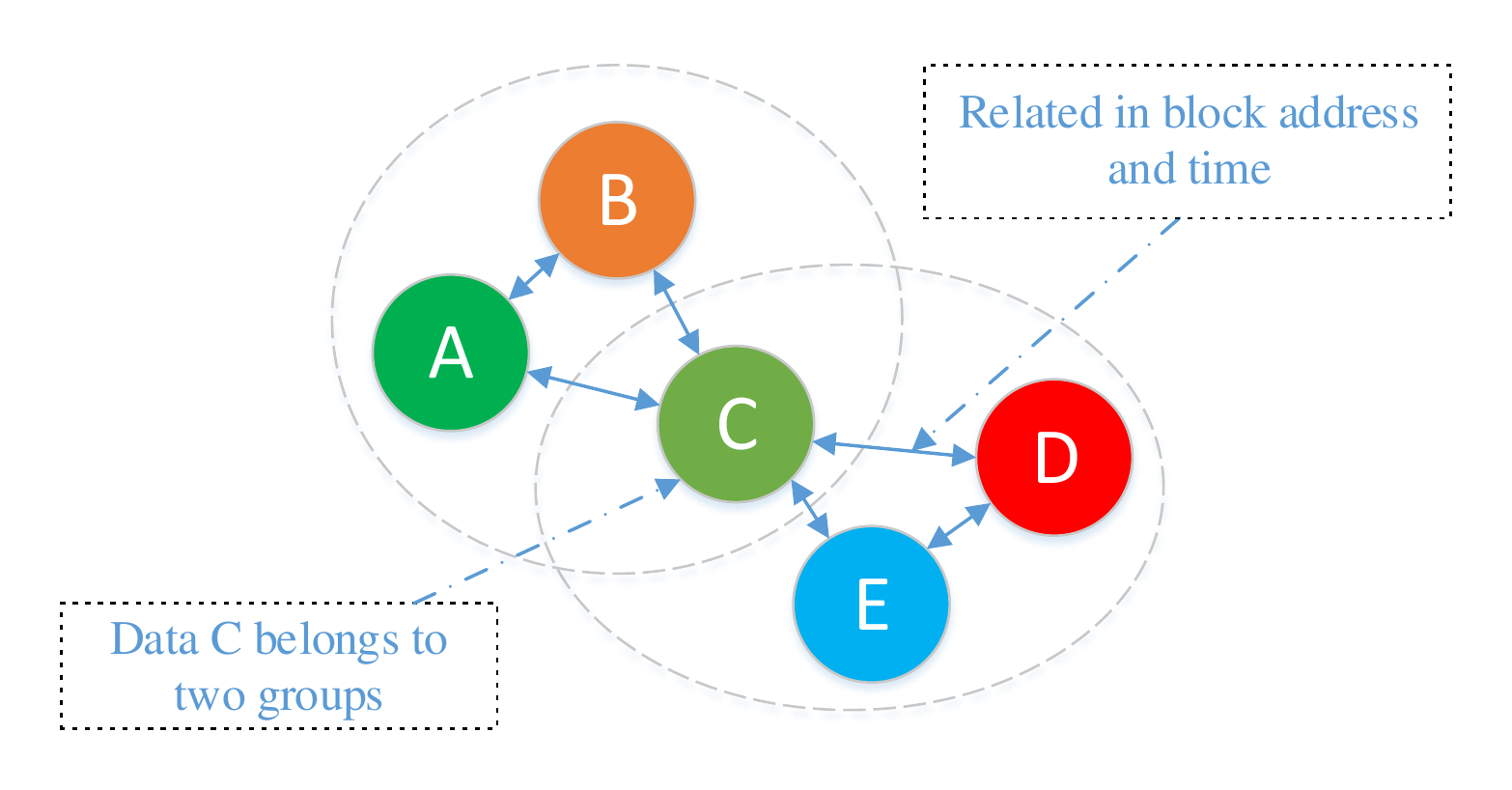}
\caption{Data C is strongly related to A/B and E/D, but the RDF-based complete subgraph search algorithm cannot provide a clear division method for C.}
\label{figure4}
\end{figure}

Existing data prefetching algorithms are mainly divided into two classes that 1) analyze association rules among data and do prefetch in units of data, and 2) divide data with strong relationships into groups, store them in contiguous locations and prefetch them in groups. We compare the popular cache prefetching algorithms in Table \ref{table_0}. For the first type of prefetching algorithm, when a data is accessed, the data related to it are prefetched and put in the cache. On this basis, researchers have proposed prefetching algorithms based on a neural network (NN)\cite{patra2010file}, frequent items\cite{banga2014proxy}, probabilistic graphing\cite{griffioen1994reducing}, and distance \cite{wildani2016can}. The total disk I/Os of this method are defined as $\textup{\textbf{len}}\left ( trace \right )*\left ( 1-accuracy \right )+\textbf{sum}\left ( prefetch \right )$. Regardless of the accuracy, the reduction of the disk I/O depends on the prefetching strategy. However, for a data access request, whether it is a cache hit or a miss, new disk access requests may be generated because all the data associated with the currently accessed data may not be included in the cache. Therefore, this method has a significant effect on the I/O response time, but it may increase the I/O burden of the disk. In the second type of prefetching algorithm, data are read in groups. When data in the group are read, the entire data group is prefetched and put in the cache. Therefore, the data in the group should be strongly correlated. The state-of-the-art graph-based data relationship mining algorithm uses the data as the point and the relationships between the pairs of data as edges to establish the relationship graph, and the complete sub-graphs are the results of the data group division\cite{wildani2016can}. This type of relationship mining algorithm comprehensively identifies data groups displaying strong correlations. However, this model cannot mine the relations between data with large block address distances since it is based on the RDF, as discussed in Section 2.2. Additionally, this model cannot provide a clear division method for data belonging to more than one group, as shown in Fig. \ref{figure4}.
  
Therefore, a data correlation mining algorithm that efficiently mines data relations with significant block address distances and divides the data into independent groups is needed.

\section{Feature extraction method based on the cache transactions}

\begin{figure*}[h]
  \centering
  \includegraphics[width=7in]{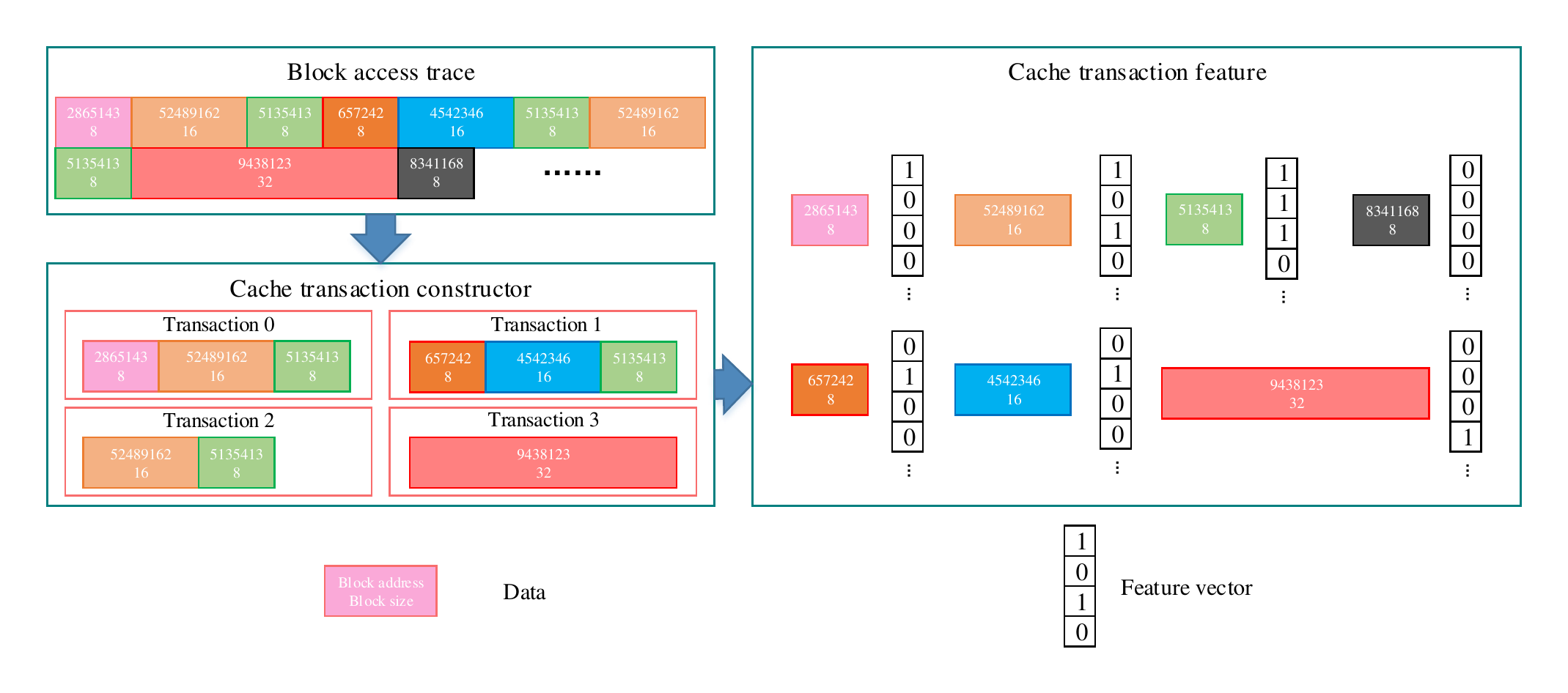}
  \caption{Schematic depicting the cache transaction feature extraction method.}
  \label{figure5}
  \end{figure*}

In this chapter, we present the definition of and method used to calculate the cache transaction, and propose a data access feature based on the cache transaction to solve the problems we discussed in Section 2. We use the size of the cache as the basis for partitioning the cache transaction to define the conditions for the access relationship between the data. The feature extraction model includes two steps: cache transaction construction and CTF extraction, as shown in Fig. \ref{figure5}.

\subsection{Definition of a cache transaction}

The cache transaction analyzes the historical data access trace and simulates a cache queue with a space size of $M$. Whenever $M$-sized data are cached, we build a cache transaction using the block address of all the data in the cache queue as the data index. Therefore, a cache transaction is understood as a collection of data indexes that may appear in the cache together.

We have attempted to incorporate multiple cache strategies as a queue for cache transactions. Finally, we selected FIFO as the strategy since it is the ideal approach in terms of both feature representation and computational efficiency. Unlike LRU, LFU, and other cache strategies, FIFO retains the timing characteristics. Although cache strategies such as LRU and LFU provide a higher cache hit rate for data accesses, these strategies cause the cache to be occupied with frequently accessed data. Thus, most of the relationships we obtain are generated between frequently accessed data, and the association between data with lower access frequencies is unable to be fully expressed. Moreover, in real storage systems, the number of data access logs is extremely large, and data access patterns change rapidly. The efficiency requirement of the feature extraction method makes the FIFO algorithm a better choice.

In this section, we provide a detailed description of the method used to calculate the cache transaction. For a cache transaction $T_i=\left \{ b_1...b_n \right \}$, $b_1...b_n$ represents the block address of data in the cache transaction. When a new data $x$ is accessed, we determine whether $b_x$ has already been added to $T_i$. If $b_x$ is included in $T_i$, then no operation is performed; otherwise, $b_x$ will be added to $T_i$ and the FIFO cache. When a cache replacement occurs, the cache replaces the size counter $Out=Out+\textup{\textbf{size}}\left ( d_{out} \right )$, where $d_{out}$ is the data that was replaced from the cache, and $\textup{\textbf{size}}$ represents a function that computes the size of a given data. When $Out \geq M$, $T_i$ will join the cache transaction queue $T$ and start the calculation of $T_{i+1}$. The pseudocode of the cache transaction division algorithm is shown in Algorithm 1.

\begin{center}
\setlength{\tabcolsep}{1mm}{ 
\begin{tabular}{{p{3.3in}}}
\toprule
\textbf{Algorithm 1}.  Cache transaction division  \\
\midrule
Input:   Max cache size $M$ \\
\quad \qquad Data access sequence $d$ \\
Output:  Cache Transactions  $T$ \\
1\,\;: \quad $T \leftarrow \left \{\right \}$\\
2\,\;: \quad $Counter \leftarrow$ 0\\
3\,\;: \quad $Out \leftarrow$ 0\\
4\,\;: \quad $Cache \leftarrow \left \{\right \}$\\
5\,\;: \quad For $i$ in $d$\\
6\,\;: \quad \qquad  if $b_i$ not in $Cache$\\
7\,\;: \quad \qquad \qquad $Cache \leftarrow Cache \cup b_i$ \\
8\,\;: \quad \qquad \qquad $Counter \leftarrow Counter + \textup{\textbf{size}}\left(i\right)$ \\
9\,\;: \quad \qquad while $Counter > M$\\
10: \quad \qquad \qquad $Counter \leftarrow Counter – \textup{\textbf{size}}\left(Cache\left[0\right]\right)$\\
11: \quad \qquad \qquad $Out \leftarrow Out + \textup{\textbf{size}}\left(Cache\left[0\right]\right)$\\
12: \quad \qquad \qquad $\textup{\textbf{Remove}}\left(Cache\left[0\right]\right)$\\
13: \quad \qquad if $Out \geq M$\\
14: \quad \qquad \qquad $Out \leftarrow$ 0\\
15: \quad \qquad \qquad $T \leftarrow T \cup$ Cache\\
16: \quad \qquad \qquad $Cache \leftarrow \left \{\right \}$\\
17: \quad Return $T$\\
\bottomrule
\end{tabular}}
\end{center}

For the size of the cache queue $M$, we generally set $M$ to less than or equal to the actual cache size in the runtime environment. A smaller $M$ ensures that the cache transaction contains fewer data than a larger $M$. Thus, data have less of a possibility to have the same feature. Therefore, $M$ is defined as the strictness of feature extraction between data.

\subsection{Data access feature representation based on the cache transaction}

\begin{figure}[h]
  \centering
  \includegraphics[width=3.5in]{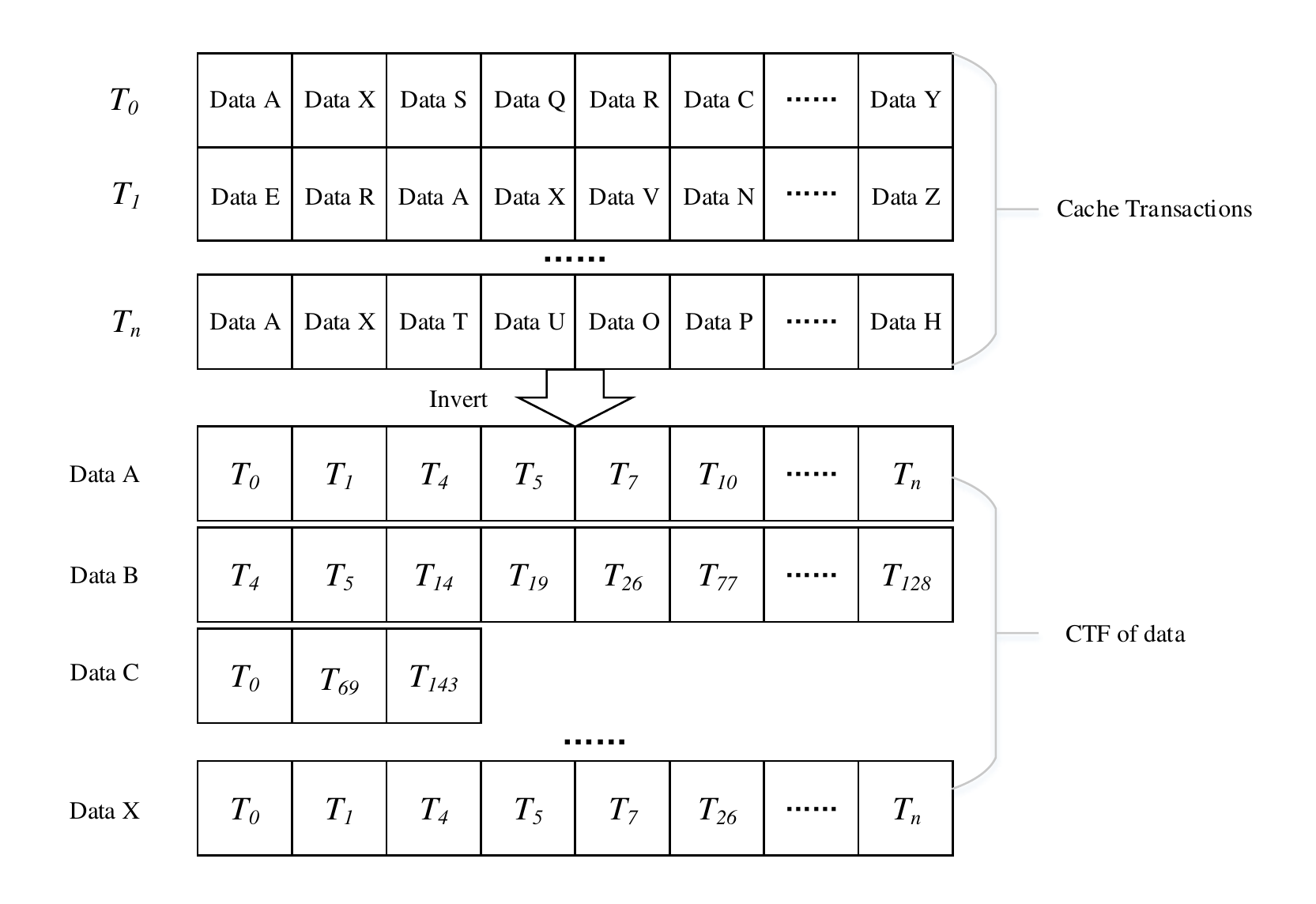}
  \caption{For cache transactions $T_{0}...T_{n}$, after inversion, we obtain the cache transaction features of all the data. We use these features as the basis for the data chunking method.}
  \label{figure6}
  \end{figure}

According to our previous analysis, a good data feature must have the ability to calculate independent features for each data. We introduce a data feature representation method based on the cache transactions that represents the appearance of each data in the cache to solve this problem. We invert the cache transactions to obtain the CTF of the data, as shown in Fig. \ref{figure6}. We design a data feature vector as $V$, where the dimension of $V$ is the total number of cache transactions. For a data $i$ and cache transaction $T_j$, if $b_i$ appears in $T_j$, then $V_{i,j}=1$; otherwise $V_{i,j}=0$, as shown in Equation (2).

\begin{equation}
V_{i,j}=\left \{{\begin{matrix} 1 & b_{i}\in T_j\\ 0 & b_{i}\notin T_j \end{matrix}} \right. \label{eq2}
\end{equation}

We obtain valuable information from this feature. For example, the Euclidean distance of two data are regarded as the access relevance of the two data. $\left| V_{i,*} \right|$ represents the data access frequency. Compared with RDF and OF, the CTF more conveniently measures the similarity between data and represents the independent feature with timing characteristics of the data. In addition, CTF is more sensitive to the size of the data in the access sequence than RDF and OF, as addressed in Section 2.1. An explanation for this finding is that larger data take up more space in a cache transaction, reducing the number of data in the cache transaction; thus, larger data are correlated with fewer data.

\section{Data grouping method}

\begin{figure*}[h]
  \centering
  \includegraphics[width=5in]{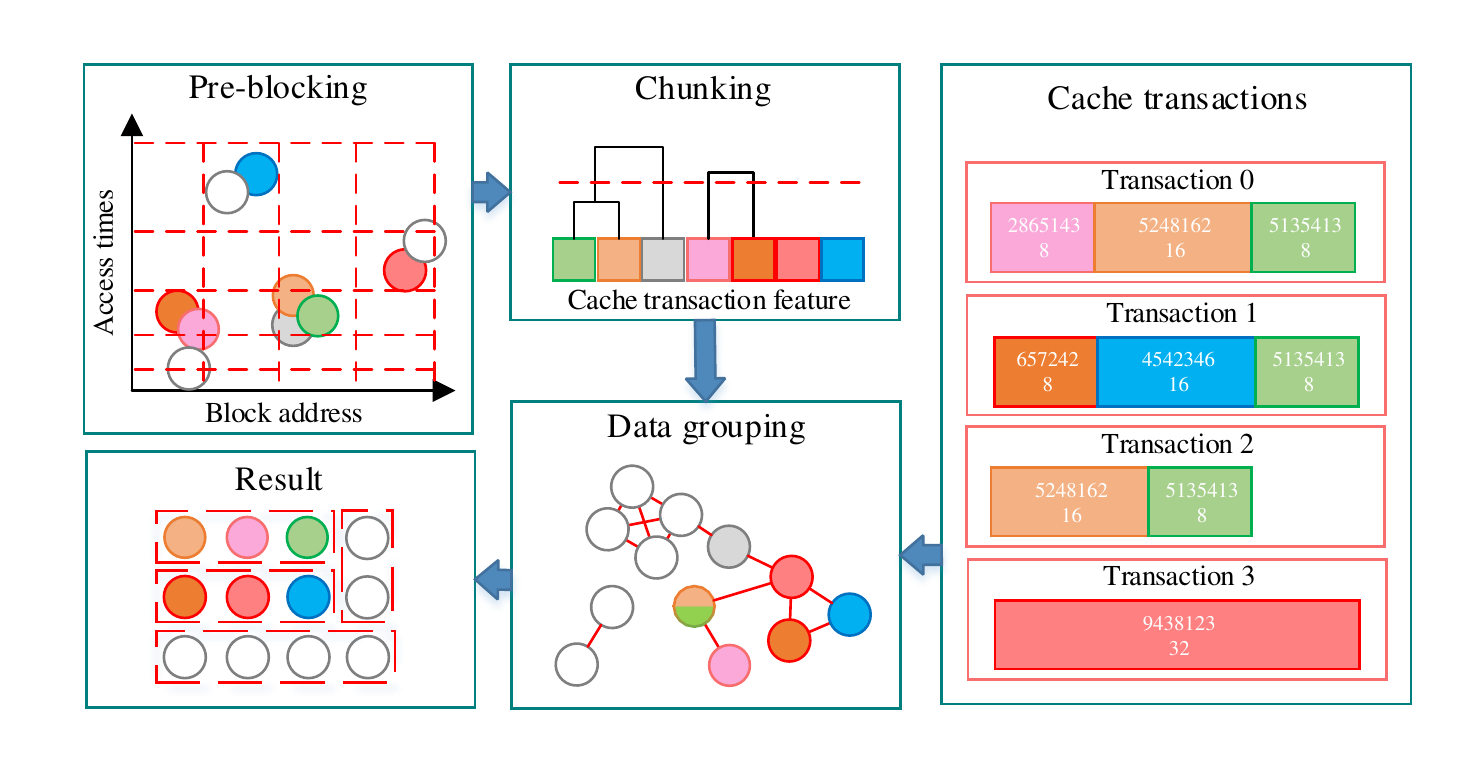}
  \caption{Schematic depicting the data grouping model. Different colored geometries represent different data. During the pre-blocking and chunking phases, the data are merged into blocks according to the block address, access times and CTF. Data, such as the green and sandy brown blocks are merged into one block. Second, we group the blocks according to the cache transactions and the relationship graph to get the final grouping result, such as the data in the dashed box in the Result process.}
  \label{figure7}
  \end{figure*}

Since CTF has advantages in terms of computability and attribute coverage, in this section, we propose a data grouping model, CTDGM, based on CTF. The model completely exploits the spatio-temporal locality features among data and strictly groups data without group-across based on the strength of the data relation. The model consists of three parts. Firstly, the data are divided into different regions according to their access frequency and block address. Secondly, the data are clustered into chunks within each region using the CTF. Finally, a data relationship graph is constructed from the relationships between the chunks and a complete subgraph searching algorithm with high correlation priority is used to obtain the data grouping result. A schematic depicting the model is shown in Fig. \ref{figure7}.

\subsection{Data chunking algorithm based on cache transaction features}

As discussed in Section 2.1, the generation of all pairwise relations among the data in a storage system is too costly. In Section 2.2, we described how closely related data display similar numbers of access times and block addresses. Therefore, we initially merge these highly correlated data into chunks. 
  
\begin{figure}[h]
  \centering
  \includegraphics[width=3.5in]{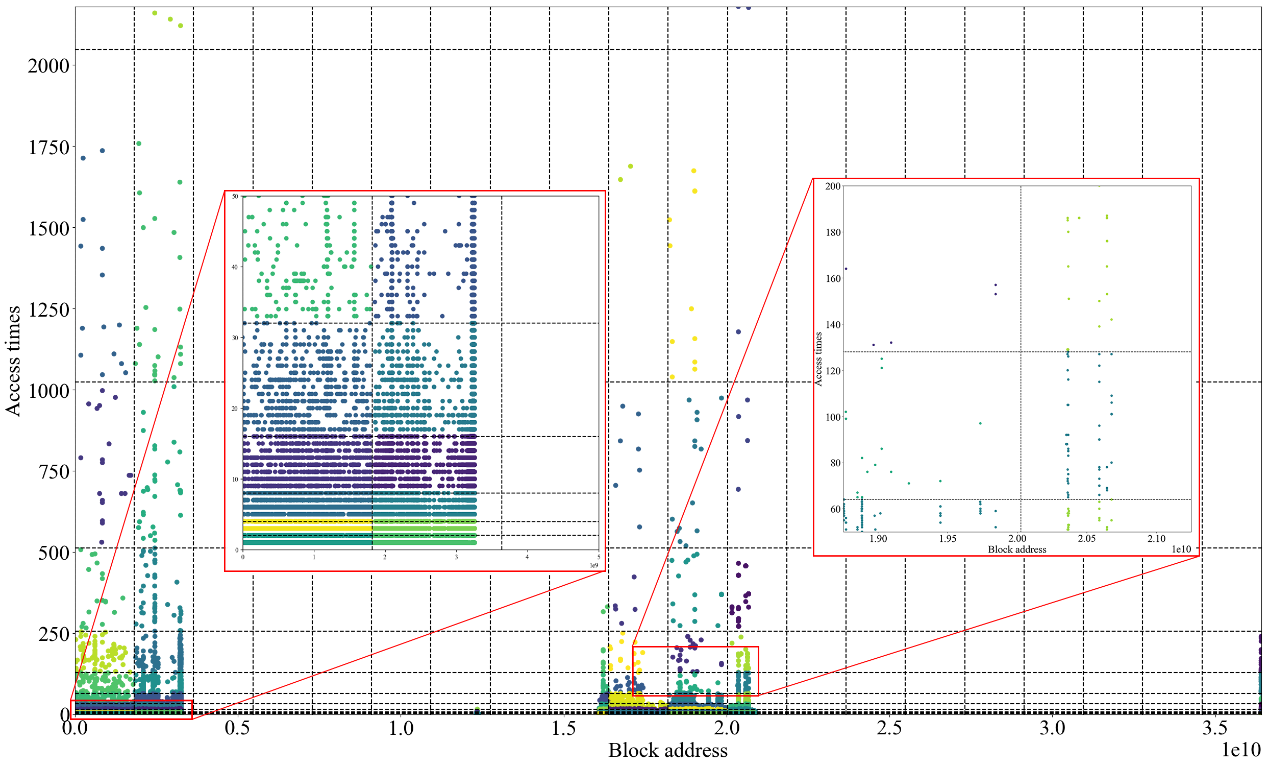}
  \caption{A sample of the data pre-blocking algorithm on MSR mds\_1 trace.}
  \label{figure8}
  \end{figure}

We propose a pre-blocking model to limit the amount of data input into the chunking algorithm and to improve the computational efficiency of the algorithm. The pre-blocking algorithm uses the access times of the data as the ordinate and the block address of the data as the abscissa to establish a plane. The plane is divided into areas $A_{i}=\left(x_{0},x_{1},y_{0},y_{1}\right)$ where  $x_0$ is the left horizontal coordinate of area $A_i$, $x_1$ is the right horizontal coordinate of the area, $y_0$ is the lower ordinate of the area $A_i$, and $y_1$ is the upper ordinate of the area. The block addresses are divided into the average-cut regions. Regarding access times, although closer access times indicate stronger relevance, the relationship is non-linear. We assume two cases for $\textup{{Data a}}$ and $\textup{{Data b}}$. In the first case, $\left | V_{\textup{Data a,*}} \right |=1$ and $\left | V_{\textup{Data b,*}} \right |=11$; in the second, $\left | V_{\textup{Data a,*}} \right |=100$ and $\left | V_{\textup{Data a,*}} \right |=110$. As we can see, in both cases, $\left | V_{\textup{Data a,*}} \right |-\left | V_{\textup{Data b,*}}\right |=10$. Where $\left | V_{\textup{Data a,*}} \right |$ is the total times $\textup{{Data a}}$ shows in all transactions. However, in the second case, a higher probability of a strong correlation exists between the two data than in the first case. Therefore, for access times, the partitioning method should be strict for data with low access times and loose for data with high access times. Assuming that the maximum block address of the storage system is $Q$ and that $Q$ is divided into $q$ regions, we number and define the regions using Equation (3):

\begin{equation}
A_{i}=\left ( \frac{i}{q}Q,\frac{i+q}{q}Q,\frac{p^{i}}{q}Q,\frac{p^{i+1}}{q}Q \right )\label{eq3}
\end{equation}

where $p$ indicates the division coefficient of the access times. A larger value of $p$ indicates a higher tolerance for the data access gap in the chunking process, and the amount of data in each area is larger than that when $p$ is small. A sample of our data pre-blocking model is shown in Fig. \ref{figure8}.

Then we then clustered the data in each area based on their CTFs. For two data, $\textup{Data }x$ and $\textup{Data }y$, the CTFs are $V_{\textup{Data}\, x,*}$ and $V_{\textup{Data}\, y,*}$. We define the relationship distance $D_{xy}$ between the data using Equation (4).

\begin{equation}
D_{xy}=\left| V_{\textup{Data}\, x,*}- V_{\textup{Data}\, y,*}\right|\label{eq4}
\end{equation}

We define a strong relationship between $\textup{Data\, }x$ and $\textup{Data\, }y$ if and only if they satisfy the relationship in Equation (5).

\begin{equation}
D_{xy}<\frac {\left| V_{\textup{Data}\, x,*}\right|+\left| V_{\textup{Data}\, y,*}\right|} {2}\sigma,\left(\sigma \leq 1\right) \label{eq5}
\end{equation}

The term $\sigma$ indicates the relationship strength threshold. When $\sigma$ is smaller, there is less tolerance for the difference in the CTF of the data. When $\sigma = 0$, only data with the same CTF are merged into data chunks. Finally, we use the data with a strong relationship $\left\{C\right\}$ as the chunks of data. The pseudocode of the data pre-blocking algorithm is shown in Algorithm 2.

\begin{center}
  \setlength{\tabcolsep}{1mm}{ 
  \begin{tabular}{p{3.3in}}
  \toprule
  \textbf{Algorithm 2}.  Data pre-blocking algorithm  \\
  \midrule
  Input:   Relation strength parameter $\sigma$ \\
  \quad \qquad Data set $DATA$ \\
  Output:  Data chunks  $C$ \\
  1\,\;: \quad $C \leftarrow \left \{\right \}$\\
  2\,\;: \quad for $d$ in $DATA$ \\
  3\,\;: \quad \qquad $k=\frac{d_{blockAddress}}{m}+mlog_{p}d_{accessTimes}$\\
  4\,\;: \quad \qquad $A_{k} \leftarrow A_{k} \cup d$ \\
  5\,\;: \quad \qquad $n=\textup{max}\left(k\right)$\\
  6\,\;: \quad for $i$ from 1 to $n$\\
  7\,\;: \quad \qquad $C \leftarrow C \cup \textup{Hcluster}\left(A_{i},\sigma\right)$ \\
  8\,\;: \quad Return $C$\\
  \bottomrule
  \end{tabular}}
  \end{center}

\subsection{Data group mining algorithm}

The data chunking algorithm described in Section 4.1 efficiently merges data with strong correlations into chunks. However, the data in a chunk belong to the same area that was divided by the pre-blocking algorithm. At this stage, we need to address only the boundary relationships across areas, which significantly reduces the number of relationships. Therefore, the data grouping model efficiently mines all data relationships, which is the bottleneck of existing algorithms. The main purpose of the data grouping algorithm is to completely mine the data relationships across area boundaries that are ignored by the data chunking algorithm, i.e., to mine the relationships between chunks. At the same time, this method effectively avoids the problem of group-across data.
  
The chunks that appear in the same cache transaction could be relevant. The condition of whether the chunk appears in a transaction is similar to that of the data. Namely, if $\textup{Data }\,x$ in chunk $C_y$ appears in $T_i$, then $C_y$ appears in $T_i$. For chunks $C_x$ and $C_y$, the access relation between them is defined as $R\left(C_x,C_y\right)$, which indicates the number of times they appeared in the same cache transaction. Since strongly related data with spatial locality have been merged into data chunks, the temporal complexity of this computational operation is nearly linear. For data with strong relations, the frequency at which they occur in the same cache transaction should be similar to the frequency at which they appear independently. Therefore, we filter the legitimacy of data relationships. We define a legal set of data relationships as $L$ and define a relationship threshold parameter $\alpha$. When the chunks satisfy Equation (6), we add a relationship between $C_x$ and $C_y$ into $L$. Subsequently, we construct the relation graph using $L$ to mine the data groups with strong correlations between chunks.

\begin{equation}
  R\left(C_x,C_y\right)\geq max\left(\left|V_{C_x}\right|,\left|V_{C_y}\right|\right)\alpha,\left(0\leq\alpha\leq1\right) \label{eq6}
  \end{equation}

We propose a complete subgraph search algorithm with a strong relation priority. Initially, we construct a data relation graph. We use the chunk as a point and the relationship between the chunks as an edge. Then, we add the relationships between chunks in $L$ to the graph according to strength from high to low. When we add $R\left(C_x,C_y\right)$ to the graph, we calculate the number of relationships for the group in which they are located using Equation (7).

\begin{equation}
  R\left(G_{C_x},G_{C_y}\right)= R\left(G_{C_x},G_{C_y}\right) + 1 \label{eq7}
  \end{equation}

Here, $G_{C_x}$ is the group to which chunk $C_x$ belongs and $R\left(G_{C_x},G_{C_y}\right)$ represents the strength of the relationship between $G_{C_x}$ and $G_{C_y}$. For the merging process, $G_{C_x}$ and $G_{C_y}$ are merged to a new complete subgraph when they satisfy the condition in Equation (8).

\begin{equation}
  R\left(G_{C_x},G_{C_y}\right)\geq \left|G_{C_x}\right| \left|G_{C_y}\right| \mu, \left(0\leq\mu\leq1\right)\label{eq8}
  \end{equation}

$\left|G_{C_x}\right|$ is the number of chunks contained in subgraph $G_{C_x}$ and $\mu$ is the relationship threshold parameter. A larger $\mu$ indicates that the data in a group have a stronger relationship with each other. The pseudocode of the data grouping model is shown in Algorithm 3.

\begin{center}
  \setlength{\tabcolsep}{1mm}{ 
  \begin{tabular}{p{3.3in}}
  \toprule
  \textbf{Algorithm 3}.  Data grouping algorithm \\
  \midrule
  Input:   Cache transaction $T$, threshold parameter $\alpha$, $\mu$\\
  Output:  Data groups  $G$ \\
  1\,\;: \quad for $t$ in $T$\\
  2\,\;: \quad \qquad $S \leftarrow \left\{\right\}$\\
  3\,\;: \quad \qquad for $d$ in $t$\\
  4\,\;: \quad \qquad \qquad $S = S \cup \textup{\textbf{Chunk}}\left(d\right)$\\
  5\,\;: \quad \qquad for $C_i,C_j$ in $S$ \\
  6\,\;: \quad \qquad \qquad $R\left(C_i,C_j\right)=R\left(C_i,C_j\right)+1$\\
  7\,\;: \quad $\textup{\textbf{Sort}}\left(R\right)$ \\
  8\,\;: \quad for $r$ in $R$ \\
  9: \quad \qquad if $R\left(r.x,r.y\right)\geq\textup{max}\left(\left|V_{r.x}\right|,\left|V_{r.y}\right|\right)\alpha$\\
  10: \quad \qquad \qquad $G_x=\textup{\textbf{Group}}\left(r.x\right)$\\
  11: \quad \qquad \qquad $G_y=\textup{\textbf{Group}}\left(r.y\right)$\\
  12: \quad \qquad \qquad if $R\left(G_x,G_y\right)\geq \left|G_{x}\right|\left|G_{y}\right|\mu$\\
  13: \quad \qquad \qquad \qquad $\textup{\textbf{Merge}}\left(G_x,G_y\right)$\\
  14: \quad \qquad \qquad \qquad $G \leftarrow G-G_{x}-G_{y}\cup G_{x,y}$\\
  15: \quad return $G$ \\
  \bottomrule
  \end{tabular}}
  \end{center}

$\textup{\textbf{Chunk}}\left(d\right)$ returns the chunk to which data $d$ belongs, $\textup{\textbf{Sort}}\left(R\right)$ sorts the relation strength into ascending order, $r.x$ and $r.y$ are the chunks of the relation $r$, and $\textup{\textbf{Group}}\left(r.x\right)$ returns the group to which chunk $r.x$ belongs.

\begin{figure}[h]
  \centering
  \includegraphics[width=3.5in]{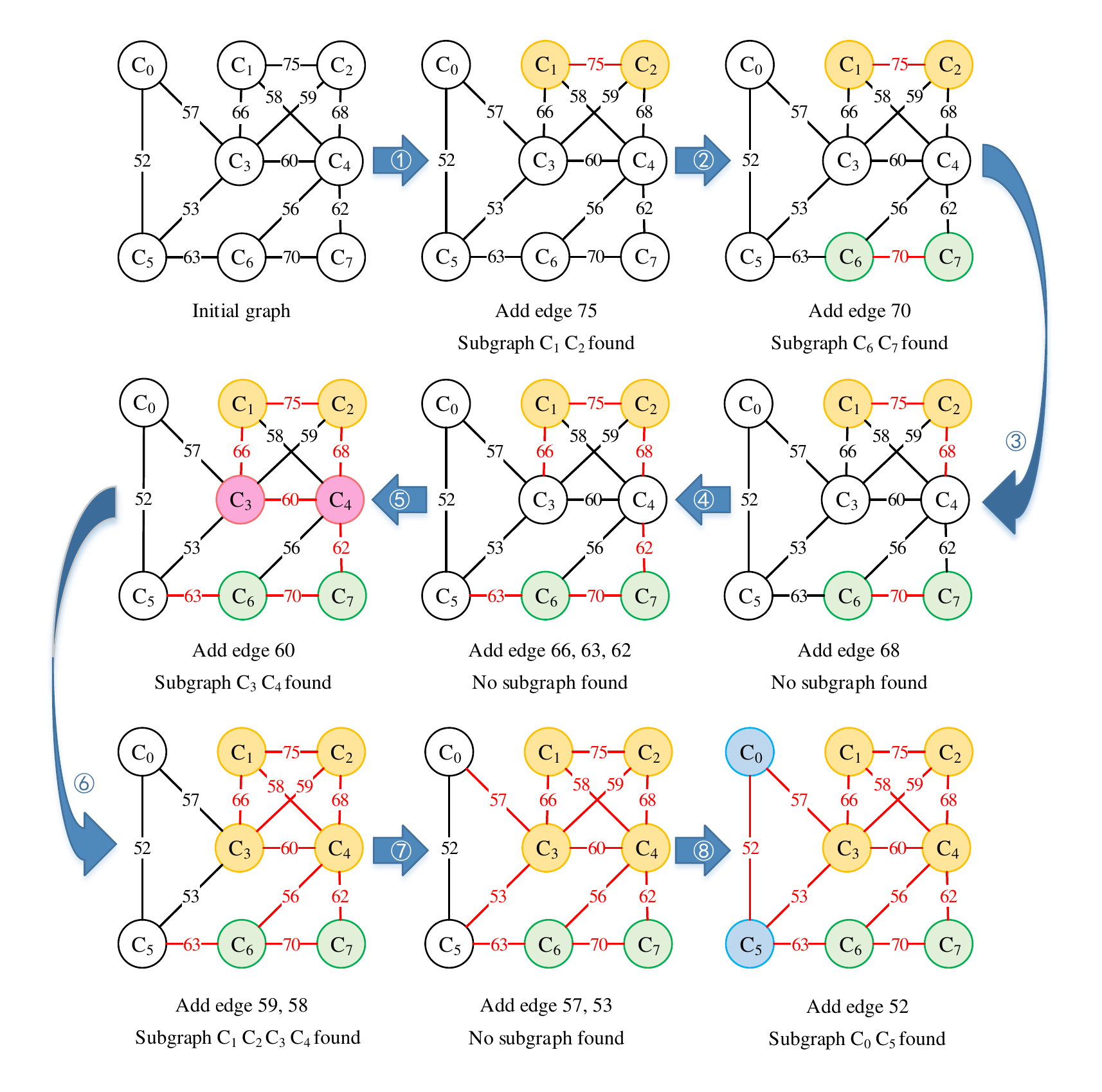}
  \caption{Schematic depicting a high-correlation priority data grouping algorithm. Nodes in different colors represent different groups, and white nodes are chunks that do not belong to any group. Black lines represent associations between groups to be mined, and red lines represent relationships between groups that have been mined.}
  \label{figure9}
  \end{figure}

Unlike the traditional complete subgraph-based data grouping model, our algorithm performs data-to-data relationship mining according to the relationship strength from high to low. Two data/groups are immediately merged into a new group, i.e., a new node in a relation graph when they satisfy Equation (8). As shown in Fig. \ref{figure9}, this method ensures that data do not simultaneously belong to multiple groups at the same time.

\section{EXPERIMENTS AND ANALYSIS}
In this chapter, we evaluate the proposed feature extraction method based on the cache transaction and data group prefetching algorithm. We compare the proposed model with popular feature extraction methods and prefetching algorithms, including prefetching algorithms based on RDF and OF, such as ANN\cite{patra2010file}, and Mithril\cite{yang2017mithril}.
  
All experiments described in this paper were performed on a Huawei RH2288 V3 server with a Xeon E5-2620 v4 CPU, 64 GB of RAM, 4 TB, 7200 RPM SATA disk, and a 4096-byte block. The system runs CentOS Linux release 7.5.1804, 64 bit, with Linux kernel Version 3.10.0. In addition, the algorithms we proposed can be deployed in a variety of distributed storage systems. We evaluate the cache hit rate performance of our model by block I/O traces. Furthermore, to show the effect on the proxy server, we implemented our model on the controller server in a three-node Openstack Swift distributed storage system to group the related data across storage nodes.
  
The dataset we selected to test our model is MSR\cite{koller2010deduplication}, which contains 1-week block I/O traces of enterprise servers on 13 servers at Microsoft Research Cambridge. MSR is widely used in storage research since it is a publicly available block I/O trace and contains multiple data access business types. Data centers run in multi-application, multi-user scenarios. These scenarios require a feature extraction method and data grouping model that can handle various types of data services. Therefore, the MSR dataset is a suitable and convincing selection for testing the performance and universal applicability of our model\cite{wu2018data}.

\begin{figure*}[h]
  \centering
  \includegraphics[width=7in,trim=110 20 120 10,clip]{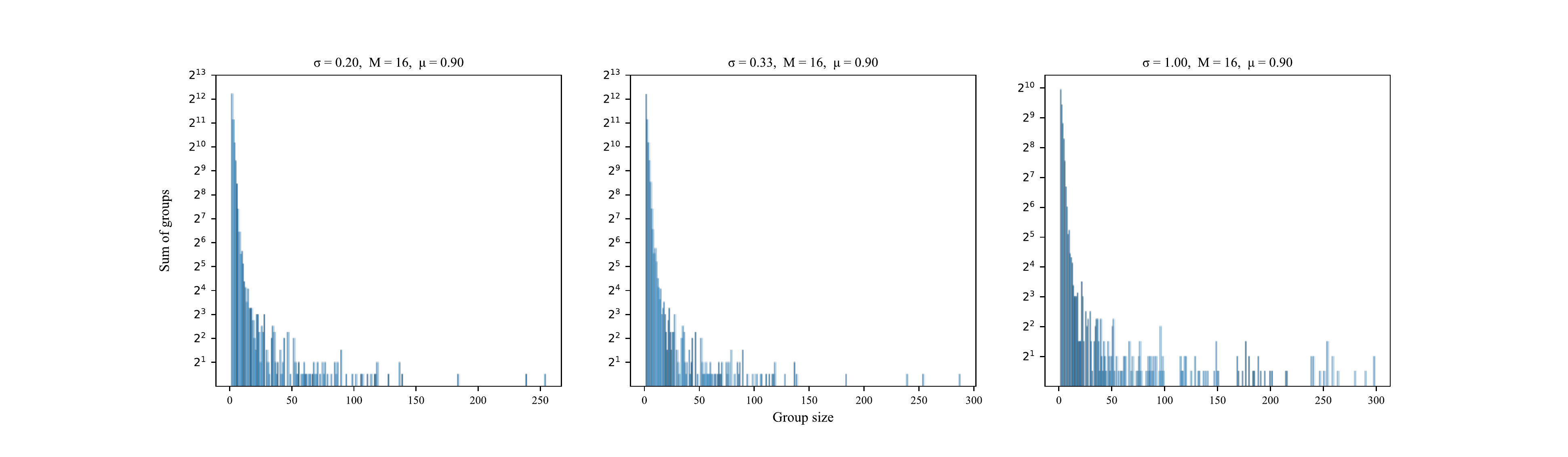}
  \caption{Effect of variations in $\sigma$ on the grouping results.}
  \label{fig_sim_10}
  \end{figure*}

The grouping algorithm adjusts the strictness of the grouping by the threshold parameter. For the cache transaction size threshold $M$ and the clustering stop threshold $\mu$, a higher threshold $\mu$ increases the strictness of the criteria for data grouping, while the number of data in each group and the total number of groups decreases. We adjusted the parameters $\sigma$ in the experiments and analyzed the changes in Fig. \ref{fig_sim_10}. As the value of parameter $\sigma$ increases, the total number of groups decreases, while each group contains more data than does a smaller $\sigma$. Thus, a more rigorous chunking process results in chunks with a finer granularity, and the correlation between the data in the chunks is stronger. This approach causes the data grouping process to work better, but it may lead to a significant increase in the computational complexity of the data grouping process.

\begin{figure*}[h]
  \centering
  \includegraphics[width=7in,trim=110 20 120 10,clip]{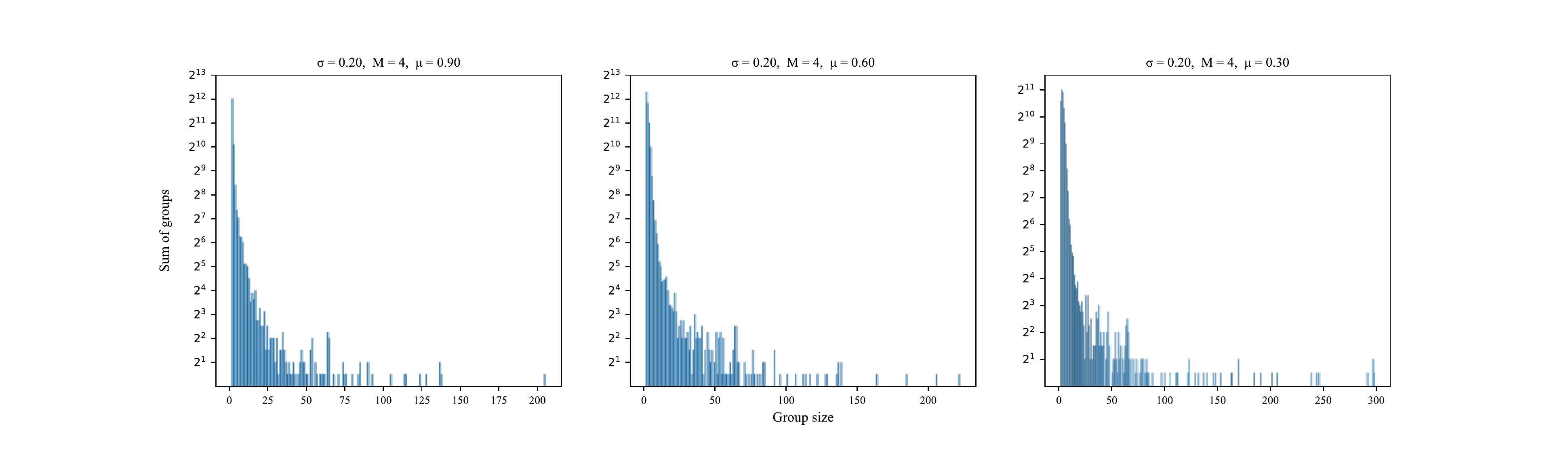}
  \caption{Effect of variations in $\mu$ on the grouping results.}
  \label{fig_sim_11}
  \end{figure*}

Additionally, as shown in Fig. \ref{fig_sim_11}, a decrease in $\mu$ weakens the conditions for merging data into groups, but it affects only the results of grouping, and the effect on computational complexity is slight. $\mu$ should be less than 1, because some of the relations across two cache transactions may be ignored by the model. The use of an excessively strict division method prevents the chunking and the grouping model from fixing these errors, and thus some of the strongly related data are unable to be merged into a group by the model.

\begin{figure*}[h]
  \centering
  \includegraphics[width=7in,trim=110 20 120 10,clip]{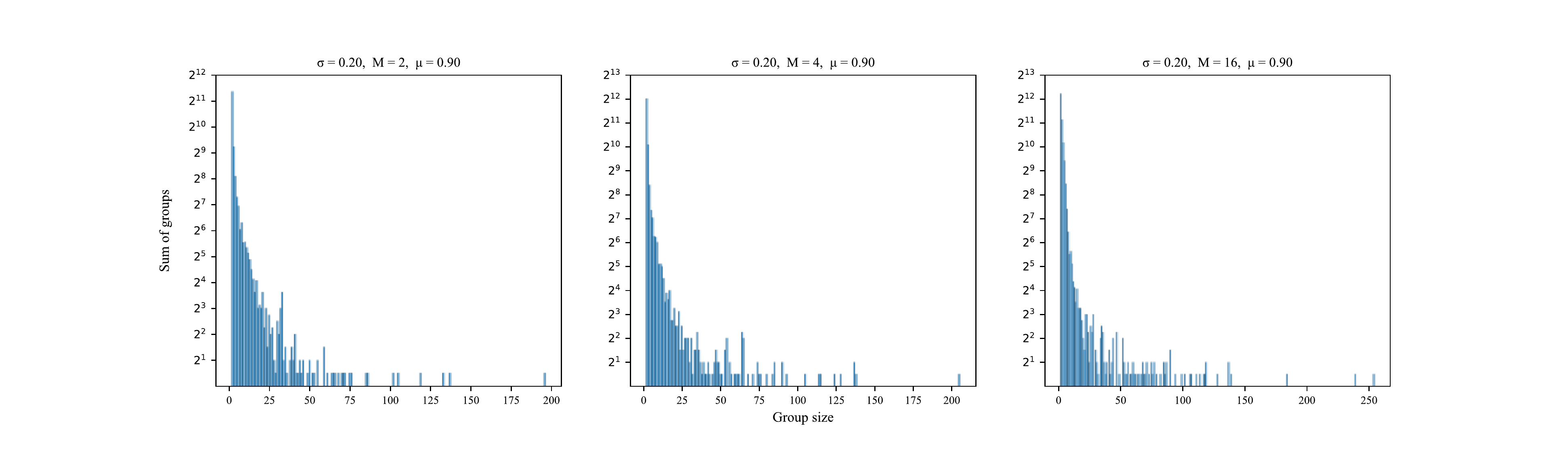}
  \caption{Effect of variations in $M$ on the grouping results.}
  \label{fig_sim_12}
  \end{figure*}

For the parameter $M$, the number of larger groups increases as $M$ increases, but the total number of groups decreases, as shown in Fig. \ref{fig_sim_12}. We analyze this parameter because the increase in the cache transaction size parameter $M$ facilitates the generation of access relationships between data. Thus, the number of data groups is easier to determine. However, a larger $M$ also potentially increases the computational complexity of the data grouping process, since the number of pairwise relationships between data is increasing. At the same time, the size of $M$ should be smaller than the actual cache size in the running environment. Otherwise, the grouping model may generate access relationships between data that do not simultaneously appear in the cache. Regarding the number of groups, the stable trends imply that the data in groups still exhibit strong relationships with each other, confirming the accuracy of the relation mining algorithm.

\begin{figure*}[h]
  \centering
  \includegraphics[width=7in,trim=130 45 130 30,clip]{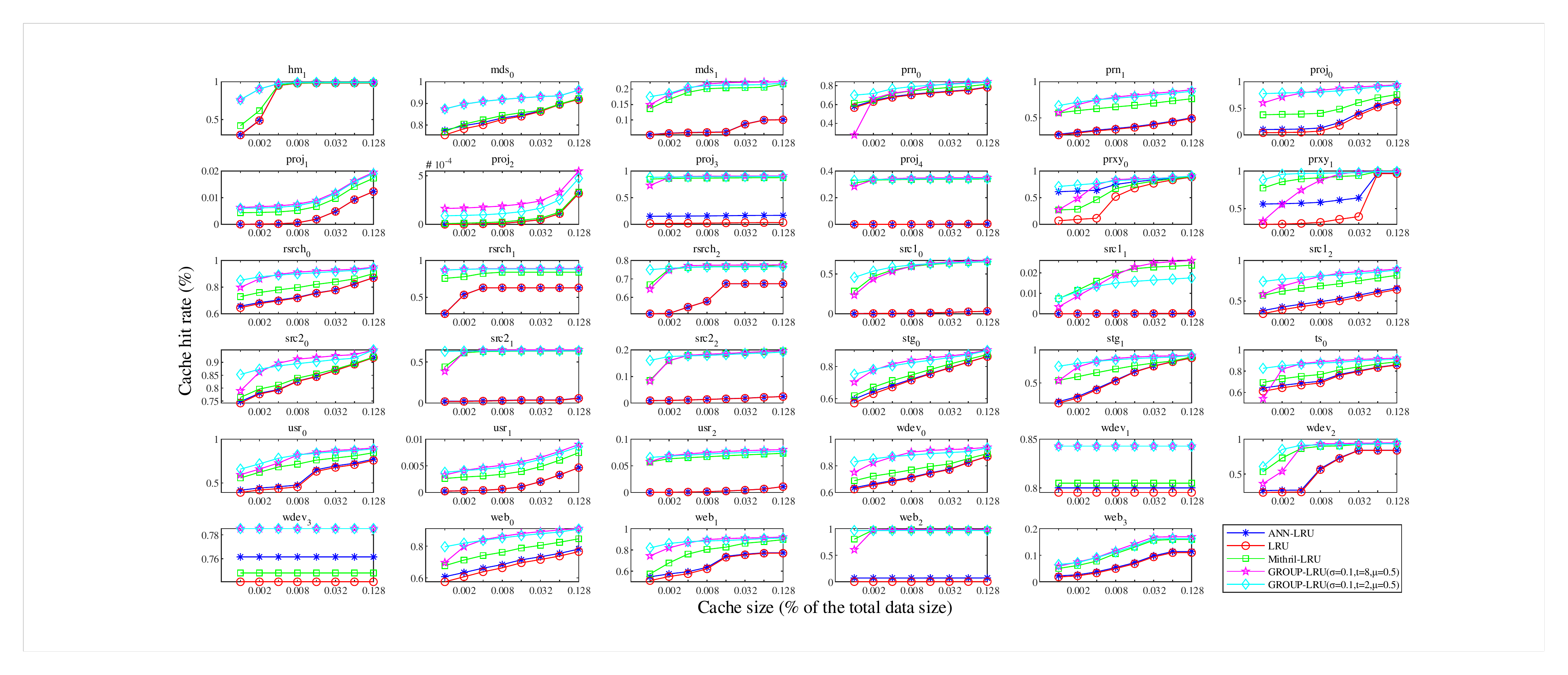}
  \caption{Cache hit rate of the MSR dataset. The cache sizes we use here are the percentage of the total data size in traces.}
  \label{figure13}
  \end{figure*}

We validated the effect of grouping on all datasets of MSR to verify the effectiveness of our data grouping algorithm. We experimented with the cache hit rate in the grouping model. Compared with the existing cache prefetching model, the model we proposed effectively improves the cache hit rate. Notably, our algorithm significantly improves the cache hit rate when the cache space is small, as shown in Fig. \ref{figure13}. 

We used three different parameters in the experiment to determine the effect of the cache transaction parameter $M$ on the grouping result. When the cache size is less than 0.0016\% and the parameter $M=8$, the effect of the grouping is not as good as when the parameter $M=2$. However, as the cache size increases, the data group with $M=8$ displays a gradually better cache hit rate than the group with $M=2$. Therefore, the mining scope of the cache transaction should be adjusted to accommodate changes in the size of the runtime environment cache.

We show the average hit rate of each algorithm in Table \ref{table_1} to more clearly display the advantages of the proposed algorithm. The CTDGM model outperforms the existing prefetching model in terms of cache hit rate, particularly when the cache space is small. Additionally, a smaller cache transaction size parameter $M$ works better in smaller runtime cache environments, which also confirms our interpretation of the data presented in Fig. \ref{fig_sim_12}.

\begin{table*}[!t]
\renewcommand{\arraystretch}{1.3}
\caption{Averange cache hit rate for prefetching models}
\label{table_1}
\centering
  \setlength{\tabcolsep}{5.3mm}{ 
  \begin{tabular}{ccccccccc}
  \toprule
  &0.001&0.002&0.004&0.008&0.016&0.032&0.064&0.128 \\
  \midrule
  LRU & 0.2908 & 0.3203 & 0.3520 & 0.3901 & 0.4287 & 0.4558 & 0.4933 & 0.5121 \\
  ANN & 0.3318 & 0.3590 & 0.3893 & 0.4174 & 0.4512 & 0.4757 & 0.5046 & 0.5223\\
  Mithril & 0.4786 & 0.5326 & 0.5733 & 0.5961 & 0.6139 & 0.6310 & 0.6471 & 0.6631\\
  Group (0.1,8,0.5) & 0.4724 & 0.5797 & 0.6326 & \textbf{0.6581} & \textbf{0.6720} & \textbf{0.6801} & \textbf{0.6857} & \textbf{0.6932}\\
  Group (0.1,2,0.5) & \textbf{0.5961} & \textbf{0.6264} & \textbf{0.6440} & 0.6545 & 0.6622 & 0.6698 & 0.6772 & 0.6869\\
  \bottomrule
  \end{tabular}}
  \end{table*}

For the system I/Os, we implemented the grouped results in two forms on proxy server in a three-node Openstack swift distributed storage system. One stores the grouping result and does not merge the data. When the storage system accesses one data, data in the same group are sequentially prefetched into the cache. In the other case, we merge the data that belong to a group and save them in a continuous storage area. When the storage system accesses one data, the entire group is prefetched into the cache. We count the number of I/Os using these two methods.

\begin{figure}
  \centering
  \subfloat[]{\includegraphics[width=3in,trim=40 20 125 30,clip]{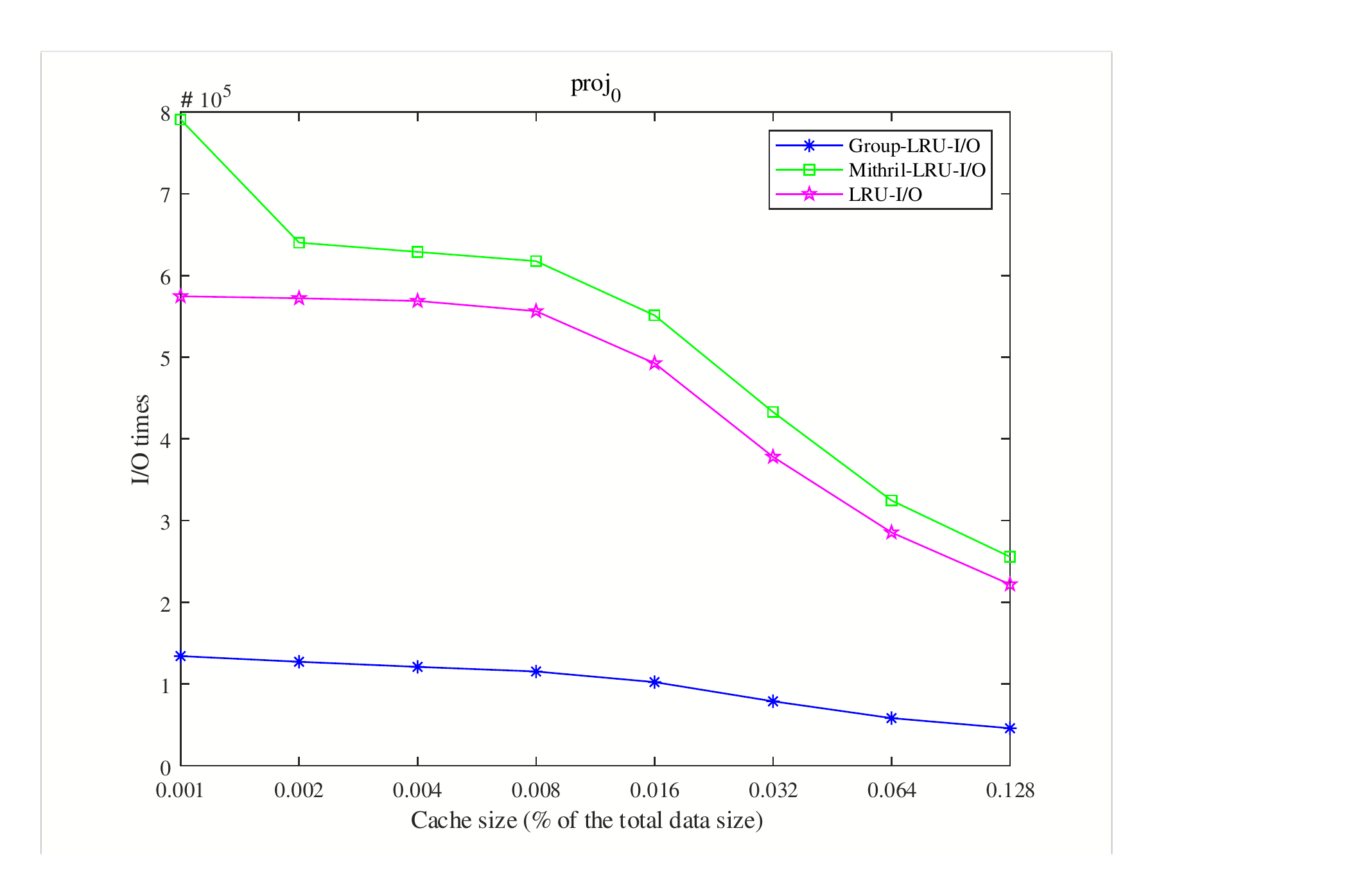}%
  \label{fig_first_case}}
  \hfil
  \subfloat[]{\includegraphics[width=3.1in,trim=40 20 125 85,clip]{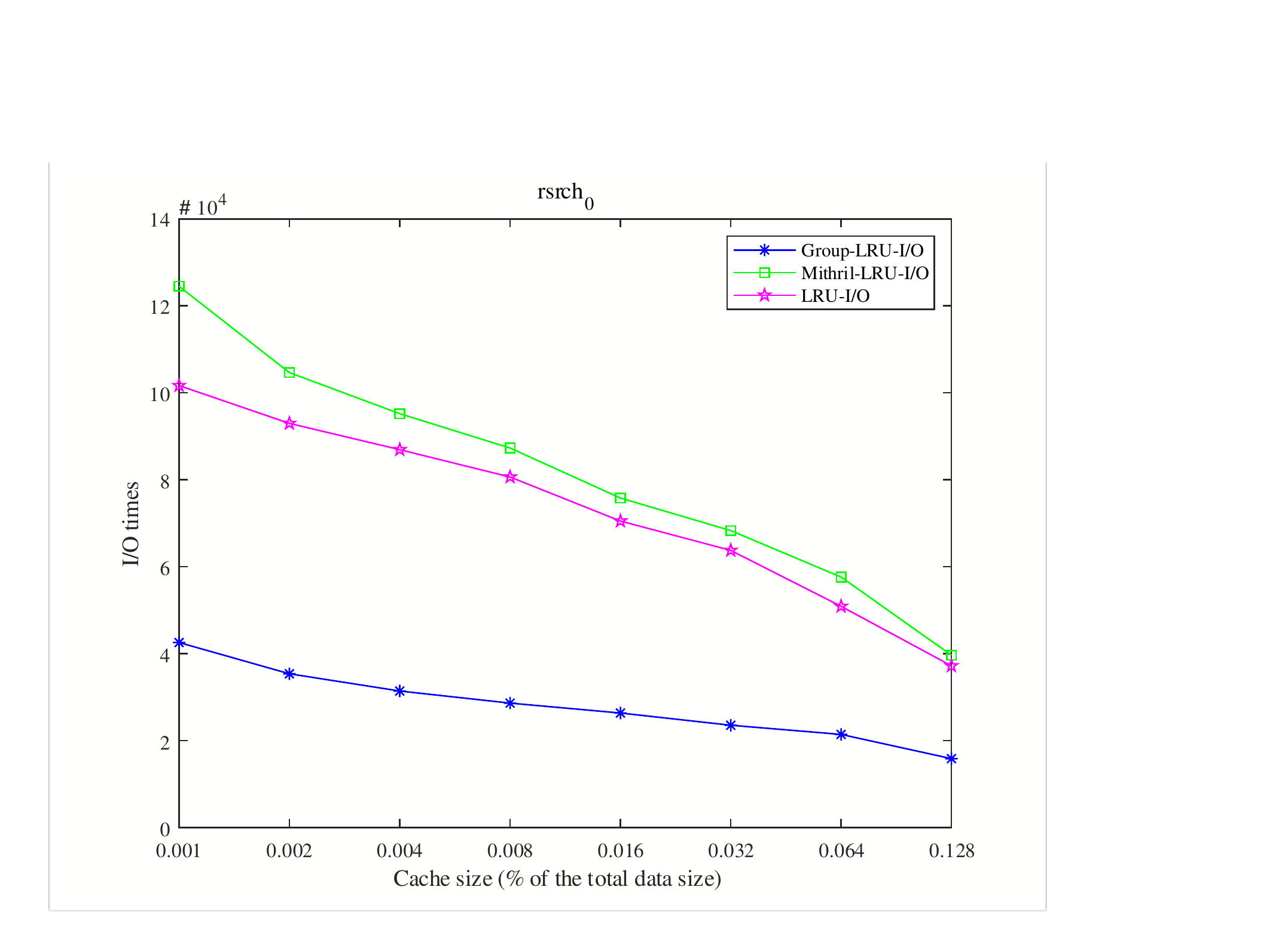}%
  \label{fig_second_case}}
  \caption{I/Os in the MSR proj\_0 and rsrch\_0 datasets}
  \label{fig_sim_14}
  \end{figure}

As shown in Fig. \ref{fig_sim_14}, when data groups are stored in continuous storage areas, the number of I/Os is up to 60\% lower than that of the one-step prefetching method possibly because the recall rate of the one-step prefetching method is relatively low since the presence of mistaken data lead to an increase in the number of I/O requests. However, I/Os decrease significantly when using the merging approach, because the related data are accessed by one I/O request. Therefore, the RDF-based data correlation mining algorithm cannot easily optimize the number of I/Os since it is unable to provide explicit group divisions of the correlated data and has a limited relation mining range.

\begin{figure}[!t]
  \centering
  \subfloat[]{\includegraphics[width=3in,trim=35 20 120 20,clip]{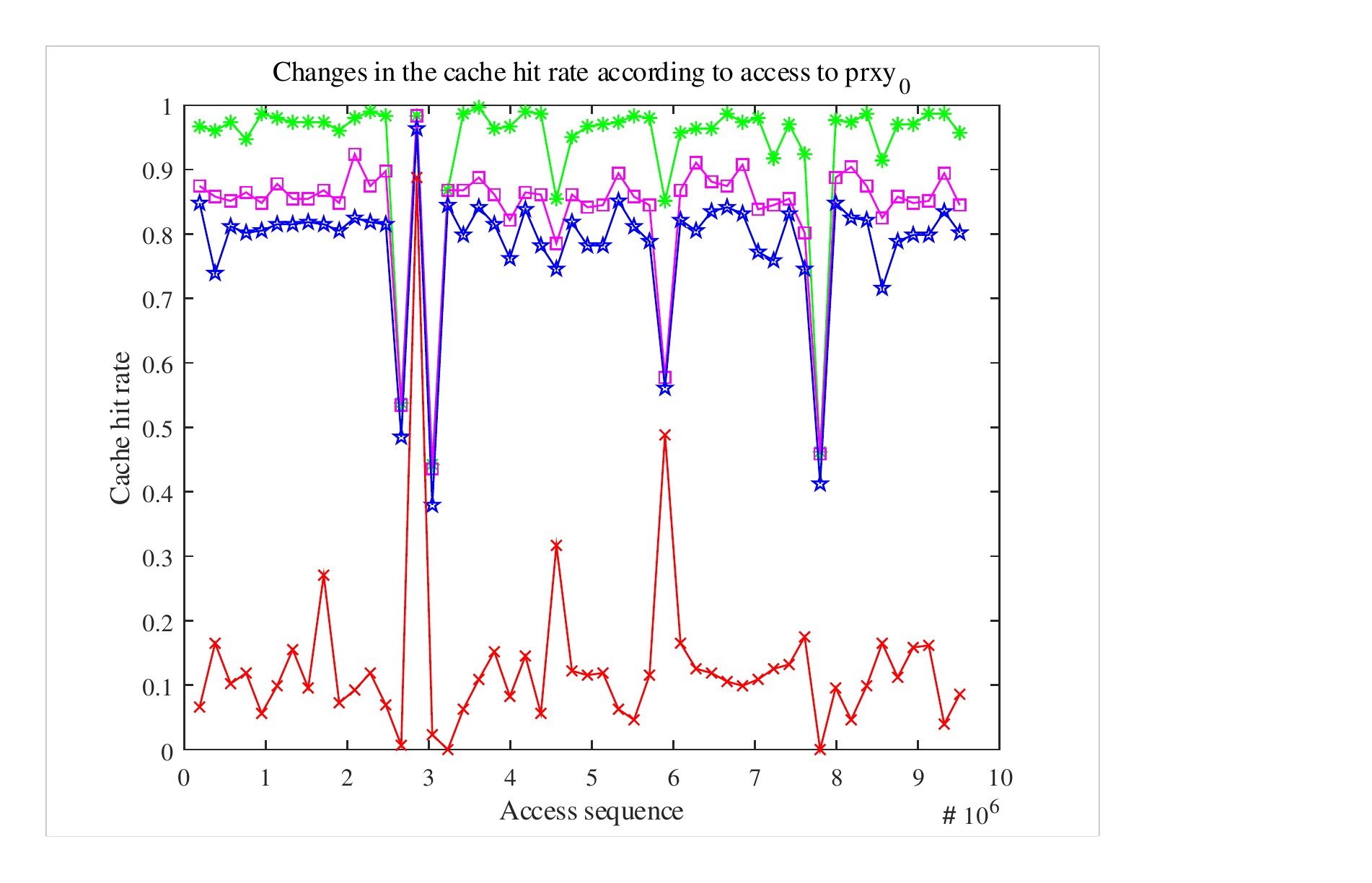}%
  \label{fig_first_case}}
  \hfil
  \subfloat[]{\includegraphics[width=2.9in,trim=40 20 140 20,clip]{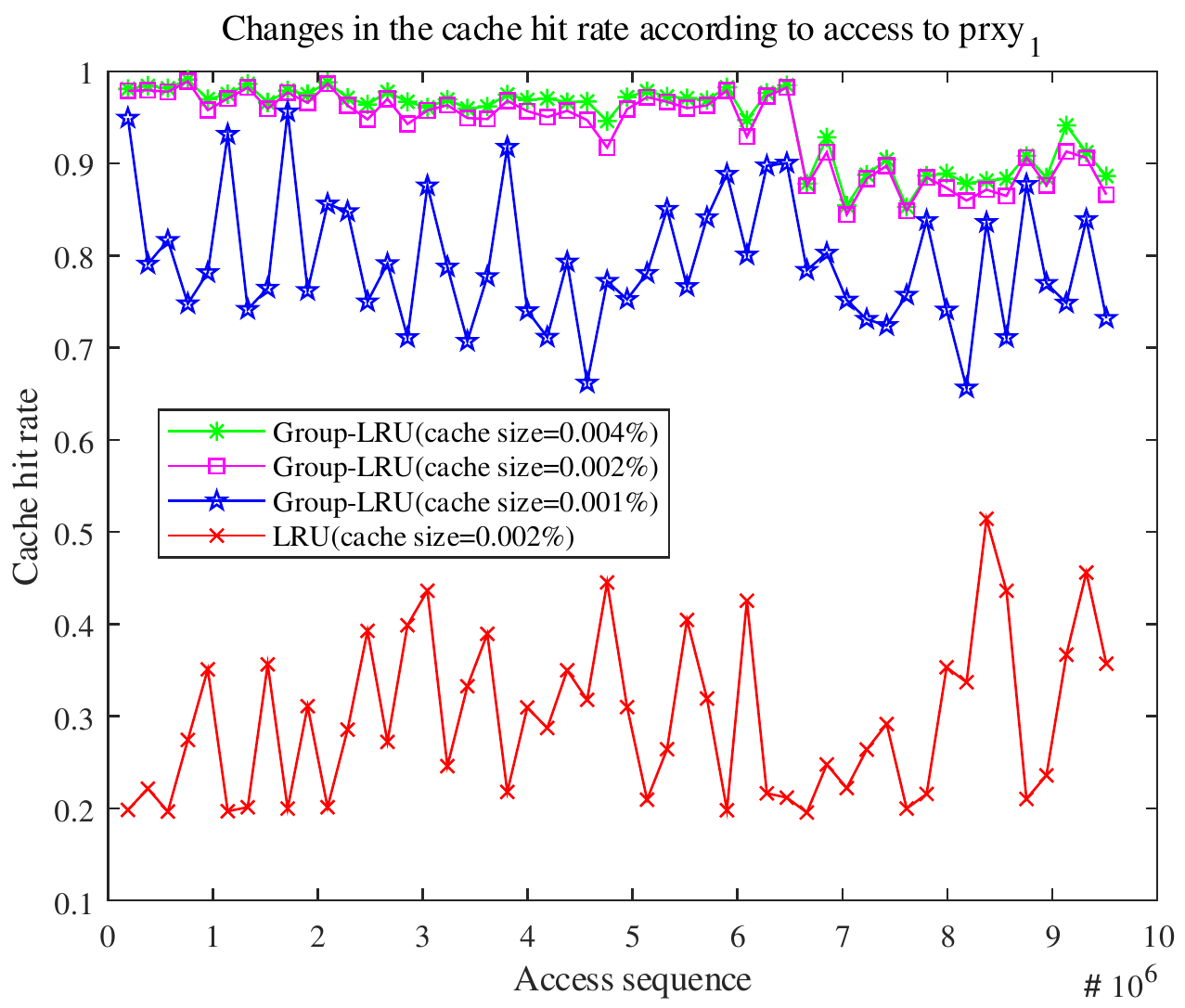}%
  \label{fig_second_case}}
  \caption{Changes in the cache hit rate over time as the number of accesses to the MSR prxy dataset increases. The cache hit rate of the LRU is used as the baseline to show the change in the access pattern.}
  \label{fig_sim_15}
  \end{figure}

As shown in Fig. \ref{fig_sim_15}, we performed a stability test on the time range of the algorithm mining results to show the stability of the relationships within the data group. The MSR\_prxy dataset has a long period and multiple users, and the user’s access interest may change. Therefore, we used this dataset to test the effect of the algorithm. In the experiment, we used the first 3 million accesses as a training set for the grouping model. Then, we used the model to verify the changes in the cache hit rate as the number of accesses changed.

For the prxy\_0 dataset, the overall trend of the cache hit rate for the proposed grouping model is similar to the LRU cache. Notably, at approximately $4.5 \times 10^6$ and $6 \times 10^6$ accesses, the cache hit rate of the grouping model decreased, and the hit rate of the LRU cache increased, possibly because the storage system has an access pattern that has not been mined by the grouping algorithm. For prefetch models, this situation is unavoidable because the model always performs data prefetching based on existing access patterns. However, its probability of occurrence is relatively low. In the prxy\_1 dataset, the performance of the cache hit rate is more stable than the that in the prxy\_0 dataset. The cache hit rate of the storage system showed an overall decrease up to approximately $6.5 \times 10^6$ accesses. We analyzed this trend because the access pattern changed. Therefore, our grouping strategy has a relatively stable effect on the cache hit rate over time; thus, we can confirm the stability of our proposed algorithm.

\section{Related  Studies}

As a method for effectively improving the access efficiency of storage systems, prefetching is a key focus area for researchers in the storage field \cite{lei1997analytical,li2008tap,gill2007amp,ding2007diskseen}. With the expansion of storage space and the development of computing power, researchers have shifted their research focus to the optimization of the accuracy and efficiency of data correlation mining algorithms. However, data have fewer feature attributes and are present at a greater quantity than in other subjects. Therefore, the main focus of data access relation mining is the representation of data correlation features and computing efficiency.

\subsection{Data access relationship characteristics}

In terms of data access feature representation, researchers initially used the OF-based approach to estimate data correlations. OF treats data as an independent object using a unique feature, such as the data storage path and one-hot encoding, to identify data. Most researchers have used the metadata and temporal locality to estimate data correlations. However, data centers have changed their storage strategies to handle the increasing amount of data. Block storage and object storage have become the primary storage strategies, making block-based data correlation mining algorithms a mainstream approach. Since blocks have fewer independent attribute features than files, researchers use an RDF to estimate data correlations. RDF combines the block address and the temporal locality of the data to define the data correlations. Thus, all properties of an RDF are numerically operated. Most studies estimate the correlations by defining the distance between data. However, the distance definition methods adopted by the RDF are mostly static and unable to easily adapt to the dynamic changes in the data access scenario. Researchers have proposed the use of the overall access density of the storage system as a parameter for dynamically defining the distance as a method to solve this problem\cite{wildani2016can}. Since the prefetch algorithm is cache-oriented, as long as two data can appear in the cache at the same time, they have a potential correlation. Additionally, the access density focuses on the level of concurrence and ignores the data size and cache size. 

We propose the cache transaction based on the content of the cache for data correlation mining, because it is not affected by changes in the data access scenario and can calculate the data correlation stably without adding parameters. Moreover, the cache transaction introduces the data size into the access feature, which expands the property coverage of the feature. We also designed the CTF based on cache transactions to ensure that the data have independent computable features. Since this feature is represented as a vector, it is also suitable for models that require vector calculation, such as a neural network. To the best of our knowledge, the CTF is the first method to establish a vector-based data-independent feature. The proposed method provides new ideas for research into data correlation mining and other fields.

\subsection{Data correlation mining algorithm}

For the data correlation mining model, the existing studies are divided into two classes. The first includes online algorithms that determine which data to prefetch when an access request arrives, such as Mithril\cite{yang2017mithril} and Tombolo\cite{yang2016tombolo}. 

Mithril prefetches data with the same access times and small access intervals as the current data. Therefore, this algorithm must establish an index of access frequency for all data stored in memory to ensure that a prefetch decision is made immediately when data are accessed. This solution quickly adapts to changes in data access patterns and displays great flexibility. However, in distributed storage systems that save massive amounts of data, Mithril requires a large amount of memory to index the data. Moreover, the scheme does not consider the effect of spatial locality on data access relevance. The calculation of the correlations among all data can place a heavy computational burden on the server. 

Tombolo uses a graph-based algorithm by establishing a one-way edge between data that are successively accessed, but it uses a substantial block address distance to form a directed graph. When data are accessed, the nodes that are reachable from the current node are the candidate set, thereby determining the prefetched data. However, the maintenance of this directed graph is costly, since the access pattern in the data center is highly random and the access frequency is high.
  
The other approach is an offline association mining algorithm that calculates the correlations among data in advance. When the data are accessed, the model prefetches the associated data and puts them in the cache according to a predetermined prefetching strategy, such as the strategies described by Wildani\cite{wildani2016can} and Zhu\cite{zhu2018access}. The method reported by Zhu calculates the access distance between data whose access times differ by less than 10\%, and the authors define the distance according to the average access interval between data. This method explicitly classifies the data but does not consider the spatial locality and the effect of the data size on the content of the cache. The method described by Wildani pre-calculates the access distance based on the block address, access time offset, and the overall data access density of the storage system. Subsequently, the algorithm filters out the data access relationships with large block address distances and uses the remaining relationship as the edge, and the data are defined as points so that a complete subgraph search can be performed. Although the subgraph searching process is dynamic, the graph is costly to maintain since it must be precalculated and saved in the memory. Additionally, the algorithm filters out most of the data that are located far from the block address. However, based on our statistical results presented in Section 2.2, existing relationships between the data have a large block address distance. Therefore, the block address distance is used to limit the mining range, which may improve the algorithm calculation speed, but decrease the accuracy of correlation mining.

\subsection{Discussion}

We propose the data correlation mining algorithm CTDGM in this paper. Unlike the existing algorithms, our algorithm estimates the access relationships of different granularities in two steps, which expands the scope of relation mining. Additionally, the algorithm is designed to clearly define data that may belong to multiple groups, which solves the across-group problem described in Section 2.3. Therefore, CTDGM still provides a high cache hit ratio when the cache space is small (0.001\% of the total data size). Thus, our model can be applied to environments other than distributed storage systems, and it even serves as a solution for disk internal caching.

We used the offline association mining algorithm because prefetched data are time-sensitive, particularly in the block layer. Although the online prefetching models may not introduce overhead in the I/O critical path since it can run in the background, the computing time for the groups may increase when the I/O concurrency is high. Therefore, the prefetched data may be out of date. However, high-concurrency scenarios are the optimal scenarios for using prefetching algorithms.
  
However, the access number covered by the CTF is limited. According to our calculations, for a 4 MB cache transaction, the CTF can cover approximately 4 TB of data access. The proposed method is superior to the existing feature extraction method in this regard, and this feature is acceptable when the access characteristics of the data are relatively stable. However, as the number of data and access concurrency in the data center continue to increase, and for complex business scenarios with multiple users and multiple applications, the data access features must cover a larger access space. We postulate that subsequent studies will be able to expand these features by optimizing the feature coding method. Additionally, CTDGM exhibits high processing efficiency and low memory consumption. In the test environment, the memory consumption of 3 million visits is less than 1 GB, and the calculation time is less than 40 seconds. Thus, our mining algorithm will continue to run on the storage node.

\section{CONCLUSION}

We propose a CTF extraction method for defining the access features of the data and a high-correlation-first data correlation mining method. Compared to existing methods, our model considers the size of the data in the correlation analysis, and the relationships between the data are clearly defined based on the cache. Additionally, using the data chunking algorithm, our method completely exploits the temporal and spatial locality features between data. It addresses the inability of the traditional method to completely exploit data relationships due to efficiency problems. Using the MSR dataset, the proposed model displays, on average, an average 12\% increase in the cache hit rate at a small cache size (0.001\% of the total data size) compared with the existing RDF and OF-based data correlation mining models. The merged group prefetching model reduces the number of I/Os by 50\% when the cache size is less than 0.008\% of the total data size, thereby increasing the concurrency of the storage system. We also propose a data feature representation method based on cache transactions, which provides a new idea for the data correlation mining algorithm.
\ifCLASSOPTIONcompsoc
  \section*{Acknowledgments}
\else
  \section*{Acknowledgment}
\fi

This work is supported by the Fundamental Research Funds for the Central Universities (grant no. HIT.NSRIF.201714), Weihai Science and Technology Development Program (2016DXGJMS15), and the Key Research and Development Program in Shandong Province (2017GGX90103).




\bibliographystyle{IEEEtran}
\bibliography{main}

\begin{thebibliography}{10}
\providecommand{\url}[1]{#1}
\csname url@samestyle\endcsname
\providecommand{\newblock}{\relax}
\providecommand{\bibinfo}[2]{#2}
\providecommand{\BIBentrySTDinterwordspacing}{\spaceskip=0pt\relax}
\providecommand{\BIBentryALTinterwordstretchfactor}{4}
\providecommand{\BIBentryALTinterwordspacing}{\spaceskip=\fontdimen2\font plus
\BIBentryALTinterwordstretchfactor\fontdimen3\font minus
  \fontdimen4\font\relax}
\providecommand{\BIBforeignlanguage}[2]{{%
\expandafter\ifx\csname l@#1\endcsname\relax
\typeout{** WARNING: IEEEtran.bst: No hyphenation pattern has been}%
\typeout{** loaded for the language `#1'. Using the pattern for}%
\typeout{** the default language instead.}%
\else
\language=\csname l@#1\endcsname
\fi
#2}}
\providecommand{\BIBdecl}{\relax}
\BIBdecl

\bibitem{shvachko2010hadoop}
K.~Shvachko, H.~Kuang, S.~Radia, R.~Chansler \emph{et~al.}, ``The hadoop
  distributed file system.'' in \emph{MSST}, vol.~10, 2010, pp. 1--10.

\bibitem{weil2006ceph}
S.~A. Weil, S.~A. Brandt, E.~L. Miller, D.~D. Long, and C.~Maltzahn, ``Ceph: A
  scalable, high-performance distributed file system,'' in \emph{Proceedings of
  the 7th symposium on Operating systems design and implementation}.\hskip 1em
  plus 0.5em minus 0.4em\relax USENIX Association, 2006, pp. 307--320.

\bibitem{bende2016dealing}
S.~Bende and R.~Shedge, ``Dealing with small files problem in hadoop
  distributed file system,'' \emph{Procedia Computer Science}, vol.~79, pp.
  1001--1012, 2016.

\bibitem{niazi2018size}
S.~Niazi, M.~Ronstr{\"o}m, S.~Haridi, and J.~Dowling, ``Size matters: Improving
  the performance of small files in hadoop,'' in \emph{Proceedings of the 19th
  International Middleware Conference}.\hskip 1em plus 0.5em minus 0.4em\relax
  ACM, 2018, pp. 26--39.

\bibitem{wang2017odds}
J.~Wang, J.~Yin, D.~Han, X.~Zhou, and C.~Jiang, ``Odds: Optimizing
  data-locality access for scientific data analysis,'' \emph{IEEE Transactions
  on Cloud Computing}, 2017.

\bibitem{chandrasekar2013novel}
S.~Chandrasekar, R.~Dakshinamurthy, P.~Seshakumar, B.~Prabavathy, and C.~Babu,
  ``A novel indexing scheme for efficient handling of small files in hadoop
  distributed file system,'' in \emph{2013 International Conference on Computer
  Communication and Informatics}.\hskip 1em plus 0.5em minus 0.4em\relax IEEE,
  2013, pp. 1--8.

\bibitem{niazi2017hopsfs}
S.~Niazi, M.~Ismail, S.~Haridi, J.~Dowling, S.~Grohsschmiedt, and
  M.~Ronstr{\"o}m, ``Hopsfs: Scaling hierarchical file system metadata using
  newsql databases,'' in \emph{15th $\{$USENIX$\}$ Conference on File and
  Storage Technologies ($\{$FAST$\}$ 17)}, 2017, pp. 89--104.

\bibitem{dong2012optimized}
B.~Dong, Q.~Zheng, F.~Tian, K.-M. Chao, R.~Ma, and R.~Anane, ``An optimized
  approach for storing and accessing small files on cloud storage,''
  \emph{Journal of Network and Computer Applications}, vol.~35, no.~6, pp.
  1847--1862, 2012.

\bibitem{lin2008amp}
L.~Lin, X.~Li, H.~Jiang, Y.~Zhu, and L.~Tian, ``Amp: an affinity-based metadata
  prefetching scheme in large-scale distributed storage systems,'' in
  \emph{2008 Eighth IEEE International Symposium on Cluster Computing and the
  Grid (CCGRID)}.\hskip 1em plus 0.5em minus 0.4em\relax IEEE, 2008, pp.
  459--466.

\bibitem{kroeger1997exploring}
T.~M. Kroeger, D.~D. Long, J.~C. Mogul \emph{et~al.}, ``Exploring the bounds of
  web latency reduction from caching and prefetching.'' in \emph{USENIX
  Symposium on Internet Technologies and Systems}, 1997, pp. 13--22.

\bibitem{dong2010novel}
B.~Dong, J.~Qiu, Q.~Zheng, X.~Zhong, J.~Li, and Y.~Li, ``A novel approach to
  improving the efficiency of storing and accessing small files on hadoop: a
  case study by powerpoint files,'' in \emph{2010 IEEE International Conference
  on Services Computing}.\hskip 1em plus 0.5em minus 0.4em\relax IEEE, 2010,
  pp. 65--72.

\bibitem{zhu2018access}
D.~Zhu, H.~Du, X.~Qiao, C.~Liu, L.~Kong, A.~Li, and Z.~Fu, ``An access
  prefetching strategy for accessing small files based on swift,''
  \emph{Procedia computer science}, vol. 131, pp. 816--824, 2018.

\bibitem{xiong2019small}
L.~Xiong, Y.~Zhong, X.~Liu, and L.~Yang, ``A small file merging strategy for
  spatiotemporal data in smart health,'' \emph{IEEE Access}, vol.~7, pp.
  14\,799--14\,806, 2019.

\bibitem{fu2014accelerating}
M.~Fu, D.~Feng, Y.~Hua, X.~He, Z.~Chen, W.~Xia, F.~Huang, and Q.~Liu,
  ``Accelerating restore and garbage collection in deduplication-based backup
  systems via exploiting historical information,'' in \emph{2014 $\{$USENIX$\}$
  Annual Technical Conference ($\{$USENIX$\}$$\{$ATC$\}$ 14)}, 2014, pp.
  181--192.

\bibitem{clements2009decentralized}
A.~T. Clements, I.~Ahmad, M.~Vilayannur, J.~Li \emph{et~al.}, ``Decentralized
  deduplication in san cluster file systems.'' in \emph{USENIX annual technical
  conference}, 2009, pp. 101--114.

\bibitem{tamersoy2014guilt}
A.~Tamersoy, K.~Roundy, and D.~H. Chau, ``Guilt by association: large scale
  malware detection by mining file-relation graphs,'' in \emph{Proceedings of
  the 20th ACM SIGKDD international conference on Knowledge discovery and data
  mining}.\hskip 1em plus 0.5em minus 0.4em\relax ACM, 2014, pp. 1524--1533.

\bibitem{li2004c}
Z.~Li, Z.~Chen, S.~M. Srinivasan, and Y.~Zhou, ``C-miner: Mining block
  correlations in storage systems.'' in \emph{FAST}, vol.~4, 2004, pp.
  173--186.

\bibitem{ge2018chewanalyzer}
X.~Ge, X.~Xie, D.~H. Du, P.~Ganesan, and D.~Hahn, ``Chewanalyzer:
  Workload-aware data management across differentiated storage pools,'' in
  \emph{2018 IEEE 26th International Symposium on Modeling, Analysis, and
  Simulation of Computer and Telecommunication Systems (MASCOTS)}.\hskip 1em
  plus 0.5em minus 0.4em\relax IEEE, 2018, pp. 94--101.

\bibitem{liao2018block}
J.~Liao, D.~Yin, and X.~Peng, ``Block i/o scheduling on storage servers of
  distributed file systems,'' \emph{Journal of Grid Computing}, vol.~16, no.~2,
  pp. 299--316, 2018.

\bibitem{wildani2016can}
A.~Wildani and E.~L. Miller, ``Can we group storage? statistical techniques to
  identify predictive groupings in storage system accesses,'' \emph{ACM
  Transactions on Storage (TOS)}, vol.~12, no.~2, p.~7, 2016.

\bibitem{yang2017mithril}
J.~Yang, R.~Karimi, T.~S{\ae}mundsson, A.~Wildani, and Y.~Vigfusson, ``Mithril:
  mining sporadic associations for cache prefetching,'' in \emph{Proceedings of
  the 2017 Symposium on Cloud Computing}.\hskip 1em plus 0.5em minus
  0.4em\relax ACM, 2017, pp. 66--79.

\bibitem{liao2015prefetching}
J.~Liao, F.~Trahay, B.~Gerofi, and Y.~Ishikawa, ``Prefetching on storage
  servers through mining access patterns on blocks,'' \emph{IEEE Transactions
  on Parallel and Distributed Systems}, vol.~27, no.~9, pp. 2698--2710, 2015.

\bibitem{liao2015performing}
J.~Liao, F.~Trahay, G.~Xiao, L.~Li, and Y.~Ishikawa, ``Performing initiative
  data prefetching in distributed file systems for cloud computing,''
  \emph{IEEE Transactions on Cloud Computing}, vol.~5, no.~3, pp. 550--562,
  2015.

\bibitem{hankins2003data}
R.~A. Hankins and J.~M. Patel, ``Data morphing: an adaptive, cache-conscious
  storage technique,'' in \emph{Proceedings of the 29th international
  conference on Very large data bases-Volume 29}.\hskip 1em plus 0.5em minus
  0.4em\relax VLDB Endowment, 2003, pp. 417--428.

\bibitem{marascu2005mining}
A.~Marascu and F.~Masseglia, ``Mining sequential patterns from temporal
  streaming data,'' in \emph{Proceedings of the first ECML/PKDD Workshop on
  Mining Spatio-Temporal Data (MSTD’05), held in conjunction with the 9th
  European Conference on Principles and Practice of Knowledge Discovery in
  Databases (PKDD’05}, 2005.

\bibitem{mishra2012discovery}
R.~Mishra and A.~Choubey, ``Discovery of frequent patterns from web log data by
  using fp-growth algorithm for web usage mining,'' \emph{International Journal
  of Advanced Research in Computer Science and Software Engineering}, vol.~2,
  no.~9, 2012.

\bibitem{jalaparti2018netco}
V.~Jalaparti, C.~Douglas, M.~Ghosh, A.~Agrawal, A.~Floratou, S.~Kandula,
  I.~Menache, J.~S. Naor, and S.~Rao, ``Netco: Cache and i/o management for
  analytics over disaggregated stores,'' in \emph{Proceedings of the ACM
  Symposium on Cloud Computing}.\hskip 1em plus 0.5em minus 0.4em\relax ACM,
  2018, pp. 186--198.

\bibitem{patra2010file}
P.~K. Patra, M.~Sahu, S.~Mohapatra, and R.~K. Samantray, ``File access
  prediction using neural networks,'' \emph{IEEE transactions on neural
  networks}, vol.~21, no.~6, pp. 869--882, 2010.

\bibitem{gill2007amp}
B.~S. Gill and L.~A.~D. Bathen, ``Amp: Adaptive multi-stream prefetching in a
  shared cache.'' in \emph{FAST}, vol.~7, no.~5, 2007, pp. 185--198.

\bibitem{yang2016tombolo}
S.~Yang, K.~Srinivasan, K.~Udayashankar, S.~Krishnan, J.~Feng, Y.~Zhang, A.~C.
  Arpaci-Dusseau, and R.~H. Arpaci-Dusseau, ``Tombolo: Performance enhancements
  for cloud storage gateways,'' in \emph{2016 32nd Symposium on Mass Storage
  Systems and Technologies (MSST)}.\hskip 1em plus 0.5em minus 0.4em\relax
  IEEE, 2016, pp. 1--14.

\bibitem{banga2014proxy}
D.~Banga and S.~Cheepurisetti, ``Proxy driven fp growth based prefetching,''
  \emph{International Journal of Advances in Engineering \& Technology},
  vol.~7, no.~3, p. 856, 2014.

\bibitem{griffioen1994reducing}
J.~Griffioen and R.~Appleton, ``Reducing file system latency using a predictive
  approach,'' in \emph{Proceedings of the USENIX Summer 1994 Technical
  Conference on USENIX Summer 1994 Technical Conference-Volume 1}.\hskip 1em
  plus 0.5em minus 0.4em\relax USENIX Association, 1994, pp. 13--13.

\bibitem{koller2010deduplication}
R.~Koller and R.~Rangaswami, ``I/o deduplication: Utilizing content similarity
  to improve i/o performance,'' \emph{ACM Transactions on Storage (TOS)},
  vol.~6, no.~3, p.~13, 2010.

\bibitem{wu2018data}
F.~Wu, B.~Zhang, Z.~Cao, H.~Wen, B.~Li, J.~Diehl, G.~Wang, and D.~H. Du, ``Data
  management design for interlaced magnetic recording,'' in \emph{10th
  $\{$USENIX$\}$ Workshop on Hot Topics in Storage and File Systems (HotStorage
  18)}, 2018.

\bibitem{lei1997analytical}
H.~Lei and D.~Duchamp, ``An analytical approach to file prefetching,'' in
  \emph{USENIX Annual Technical Conference}, 1997, pp. 275--288.

\bibitem{li2008tap}
M.~Li, E.~Varki, S.~Bhatia, and A.~Merchant, ``Tap: Table-based prefetching for
  storage caches.'' in \emph{FAST}, vol.~8, 2008, pp. 1--16.

\bibitem{ding2007diskseen}
X.~Ding, S.~Jiang, F.~Chen, K.~Davis, and X.~Zhang, ``Diskseen: Exploiting disk
  layout and access history to enhance i/o prefetch.'' in \emph{USENIX Annual
  Technical Conference}, vol.~7, 2007, pp. 261--274.

\end{thebibliography}
%



%

\begin{IEEEbiography}[{\includegraphics[width=1in,height=1.25in,clip,keepaspectratio]{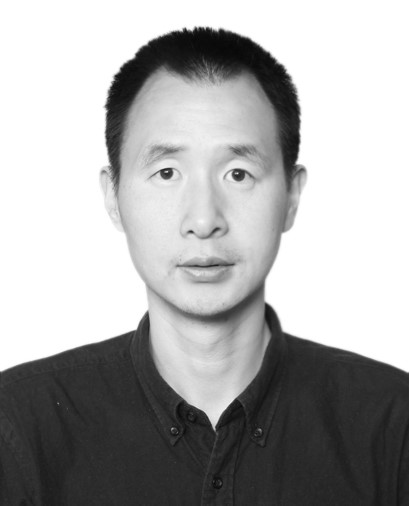}}]{Dongjie Zhu}
  received the PhD degree in computer architecture from the Harbin Institute of Technology, in 2015. He is a assistant professor in School of Computer Science and Technology at Harbin Institute of Technology, Wehai. His research interests include parallel storage systems, social computing.
\end{IEEEbiography}

\begin{IEEEbiography}[{\includegraphics[width=1in,height=1.25in,clip,keepaspectratio]{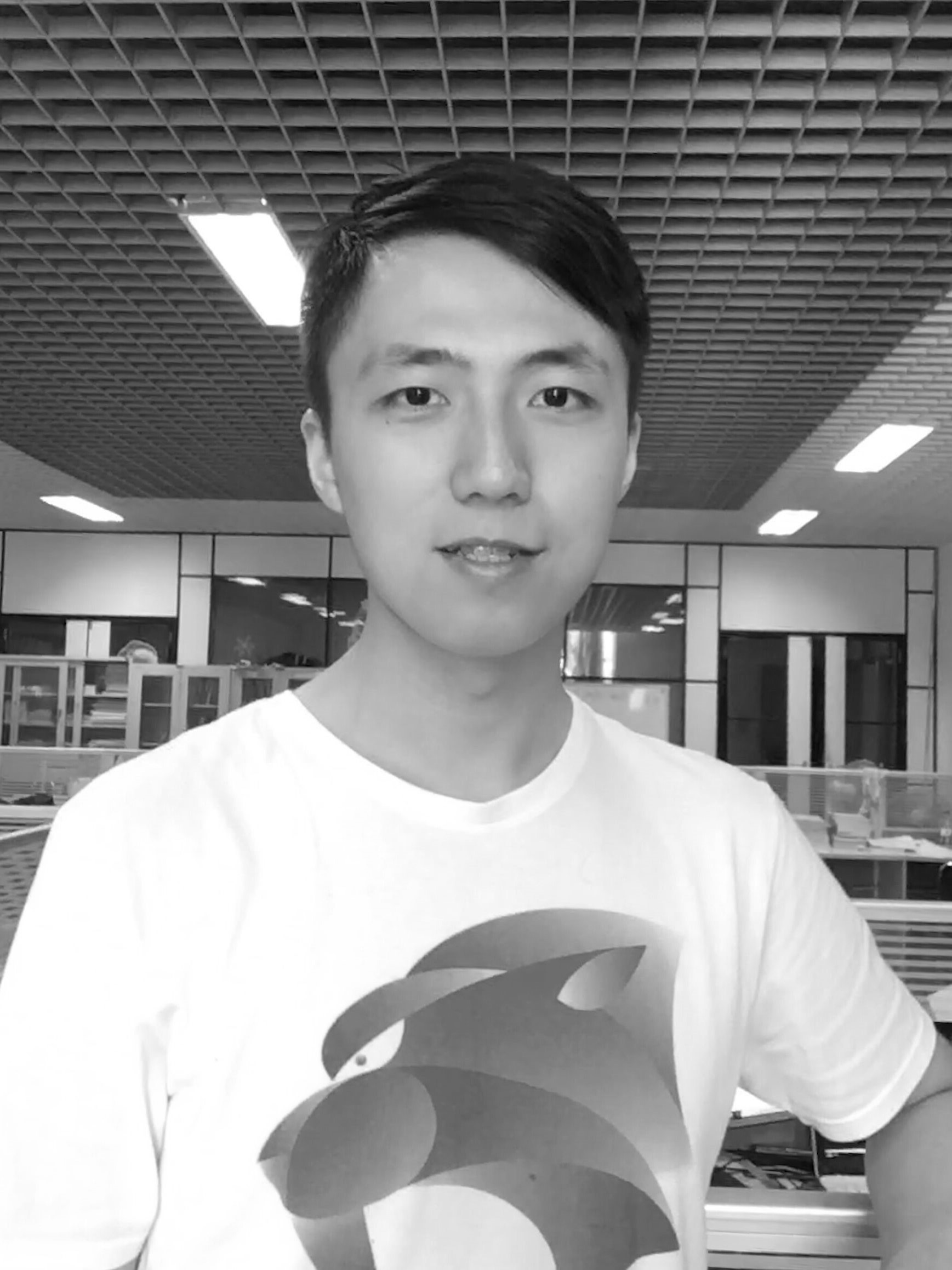}}]{Haiwen Du}
  is currently working toward the PhD degree in School of Astronautics at Harbin Institute of Technology. His research interests include storage system architecture and massive data management.
  \end{IEEEbiography}
  
\begin{IEEEbiography}[{\includegraphics[width=1in,height=1.25in,clip,keepaspectratio]{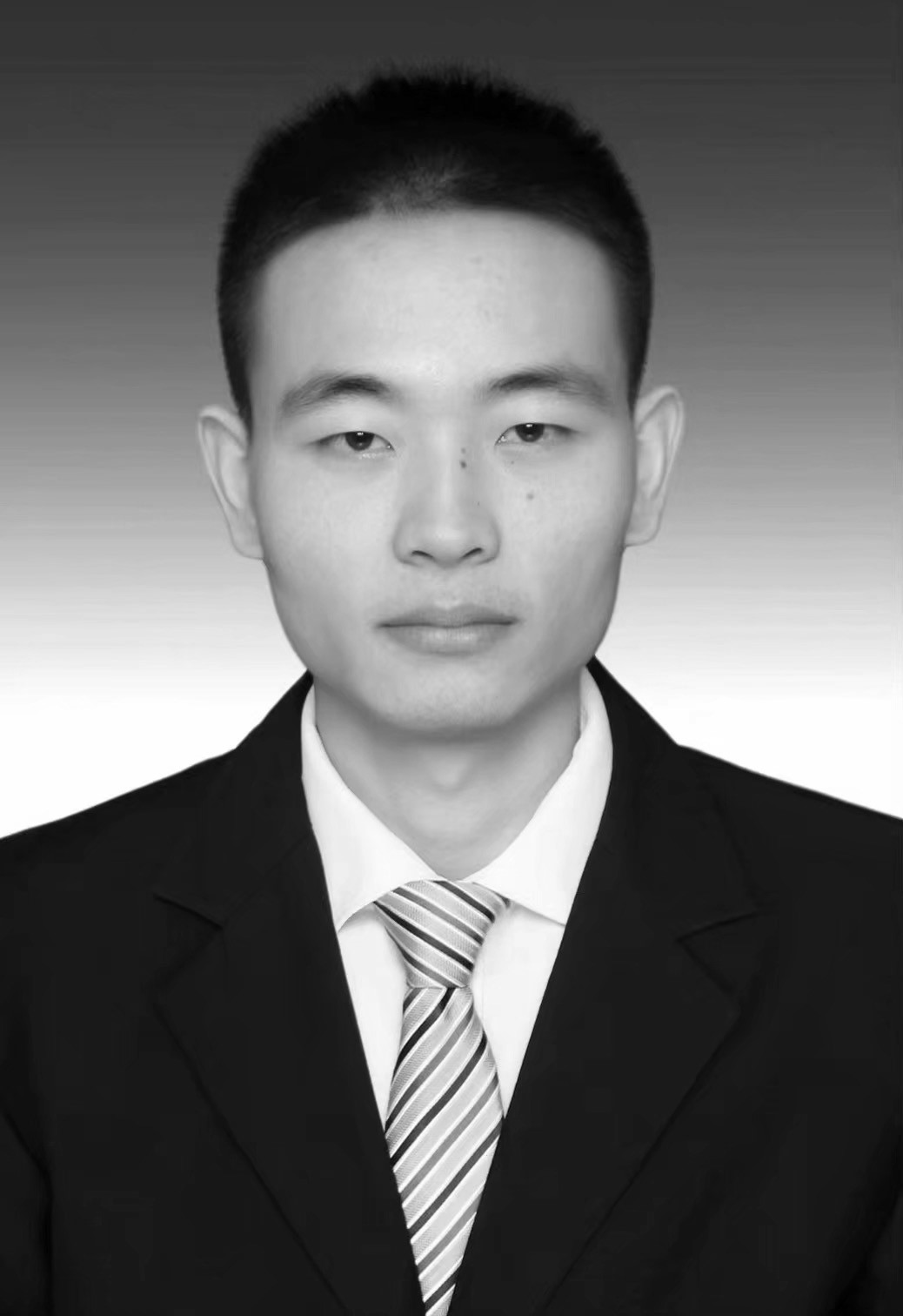}}]{Yundong Sun}
  is currently working toward the Master degree in School of Computer Science and Technology at Harbin Institute of Technology. His research interests include social computing and network embedding.
  \end{IEEEbiography}
  
\begin{IEEEbiography}[{\includegraphics[width=1in,height=1.25in,clip,keepaspectratio]{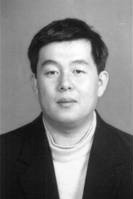}}]{Zhaoshuo Tian}
  is a professor in School of Astronautics at Harbin Institute of Technology. His research interests include laser technology and marine laser detection technology. He is a member of IEEE.
  \end{IEEEbiography}






\end{document}